\documentclass[12pt,a4]{article} 
\newcommand{\vs}{\vspace{-0.25cm}}
\usepackage{amsmath,graphicx}
\begin{document} 
\begin{center}
  {\Large{\bf Density-dependent NN-interaction from subleading chiral
three-nucleon forces: Long-range terms}\footnote{This work 
has been supported in part by DFG and NSFC (CRC110).}  }  

\medskip

 N. Kaiser$^1$ and B. Singh$^{1,2}$ \\
\medskip
{\small $^1$ Physik-Department T39, Technische Universit\"{a}t M\"{u}nchen,
   D-85747 Garching, Germany\\ $^2$ Indian Institute of Science Education and Research, Bhopal 462 066, Madhya Pradesh, India }
\end{center}
\medskip
\begin{abstract}
We derive from the subleading contributions to the chiral three-nucleon force
(long-range terms, published in Phys.\,Rev.\,C\,77, 064004 (2008)) a
density-dependent two-nucleon interaction $V_\text{med}$ in isospin-symmetric, spin-saturated nuclear matter. Following the division of the pertinent 3N-diagrams into two-pion exchange topology, two-pion-one-pion exchange topology and ring topology, we evaluate for these all self-closings and concatenations of nucleon-lines to an in-medium loop. The momentum and $k_f$-dependent potentials associated with the isospin operators ($1$ and $\vec\tau_1\!\cdot\!\vec\tau_2$) and five
independent spin-structures are expressed in terms of functions, which are
either given in closed analytical form or require at most one numerical
integration. In the same way we treat the $2\pi$-exchange 3N-force at N$^4$LO. Our results for $V_\text{med}$ are most helpful to implement the
long-range subleading chiral 3N-forces into nuclear many-body calculations.
\end{abstract}

\section{Introduction and summary}
Three-nucleon forces are an indispensable ingredient in accurate few-nucleon and
nuclear structure calculations. Nowadays, chiral effective field theory is the
appropriate tool to construct systematically the nuclear interactions in harmony with the symmetries of QCD. Three-nucleon forces appear first at N$^2$LO, where they consist of a zero-range contact-term ($\sim c_E$), a mid-range $1\pi$-exchange component ($\sim c_D$) and a long-range $2\pi$-exchange component ($\sim c_{1,3,4}$). The calculation of the subleading chiral three-nucleon forces, built up by many pion-loop diagrams, has been performed for the long-range contributions in ref.\cite{3Nlong} and completed with the short-range terms and relativistic $1/M$-corrections in ref.\,\cite{3Nshort}. Moreover, the extension of the chiral three-nucleon force to sub-subleading order (N$^4$LO) has been acomplished  for the longest-range $2\pi$-exchange component in ref.\,\cite{twopi4} and for the intermediate-range contributions in ref.\,\cite{midrange4}. Very recently, the $2\pi$-exchange component of the 3N-force has also been analyzed in chiral effective field theory with $\Delta(1232)$-isobars as explicit degrees of freedom \cite{twopidelta} at order N$^3$LO.

However, for the variety of existing many-body methods, that are commonly employed in calculations of nuclear matter or medium mass and heavy nuclei, it is technically very challenging to include the chiral three-nucleon forces directly. An alternative and simpler approach is to use instead a density-dependent
two-nucleon interaction $V_\text{med}$ that reflects the underlying three-nucleon force. The analytical calculation of $V_\text{med}$ from the leading chiral 3N-force at N$^2$LO (involving the parameters $c_{1,3,4}$, $c_D$ and $c_E$) has been presented in ref.\,\cite{holt}. When restricting to on-shell scattering of two nucleons in isospin-symmetric spin-saturated nuclear matter, the resulting in-medium NN-potential  $V_\text{med}$ has the same isospin- and spin-structure as the free NN-potential. A subsequent decomposition into partial-wave matrix elements has provided a good illustration of the (repulsive or attractive) effects of the various components of $V_\text{med}$ in different spin-isospin and angular momentum channels. The in-medium NN-interaction $V_\text{med}$ as derived from the leading chiral 3N-force has actually found many applications in recent years, e.g. for studying the thermodynamic properties of nuclear matter \cite{corbinian1,corbinian2}, or for calculations of cold nuclear and neutron matter up to third order in many-body perturbation theory \cite{holt3} and spin-polarized neutron matter \cite{samma}.  The normal ordering of
the subleading chiral 3N-forces to density-dependent NN-interactions has been performed previously by the Darmstadt group using a decomposition in a $Jj$-coupled 3N partial-wave momentum basis and applied in second order many-body
perturbation theory to the equation of state of isospin-asymmetric nuclear
matter \cite{normalorder}. In the same many-body approach, the energy per
particle of pure neutron matter \cite{achim1}, the $nn$-pairing gaps in the
$^1\!S_0$ and coupled $^3\!P_2$-$^3\!F_2$ channels \cite{achim2}, and the
saturation properties of isospin-symmetric nuclear matter  \cite{achim3} have
been studied extensively. Note that in this approach the treatment of the
chiral 3N-forces happens entirely in numerical form.

\begin{figure}[h]
\centering
\includegraphics[width=0.8\textwidth]{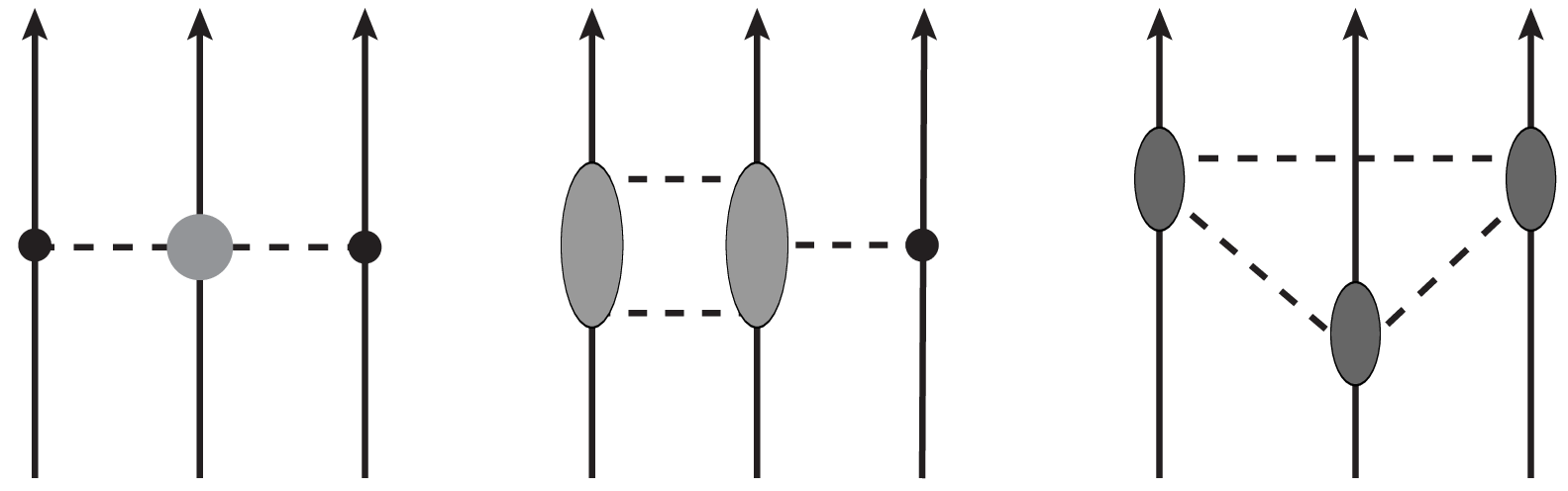}
\caption{$2\pi$-exchange, $2\pi1\pi$-exchange and ring topologies that appear in the long-range chiral 3N-interaction at subleading order.} \end{figure}

In a recent work \cite{vmedshort} we have derived from the subleading contributions to the chiral three-nucleon force (short-range terms and relativistic $1/M$-corrections \cite{3Nshort}) the density-dependent NN-interaction $V_\text{med}$ in isospin-symmetric nuclear matter of density $\rho=2k_f^3/3\pi^2$. The momentum and $k_f$-dependent potentials associated with the isospin-operators ($1$ and $\vec\tau_1\!\cdot\!\vec\tau_2$) and five independent spin-structures could be expressed in terms of loop-functions, which were either given in closed analytical form or required at most one numerical integration. The purpose of the present paper is to perform the analogous but more demanding calculation for the long-range terms of the subleading chiral 3N-force, which were divided in ref.\,\cite{3Nlong} into diagram classes of two-pion exchange topology, two-pion-one-pion exchange topology and ring topology. These topologies are depicted in Fig.\,1, where the solid dot denotes the ordinary pion-nucleon coupling, the circular blob represents pion-loops and the oval-shaped blobs include different orderings and types of pion-nucleon vertices.  Following previous works \cite{holt,vmedshort}, the in-medium NN-interaction $V_\text{med}$ is derived from the on-shell scattering process $N_1(\vec p\,)+ N_2(-\vec p\,) \to N_1(\vec p\,')+N_2(-\vec p\,')$ in the center-of-mass frame, which coincides with the nuclear matter rest frame. The ingoing and outgoing momenta are $\pm \vec p$ and $\pm \vec p\,'$, where $|\vec p\,|=|\vec p\,'|=p$ holds, and $\vec q=\vec p\,'-\vec p$ is the momentum transfer with $0\leq q\leq2p$. For NN-scattering in 
isospin-symmetric spin-saturated nuclear matter of density $\rho=2k_f^3/3\pi^3$ the isospin operators $1$ and $ \vec\tau_1\!\cdot\! \vec\tau_2$ occur, and both parts of $V_\text{med}$ can be expanded in terms of the five independent spin operators: $1\,,  \vec\sigma_1\!\cdot\!\vec\sigma_2\,,\vec\sigma_1\!\cdot\!\vec q\, \vec\sigma_2\!\cdot\! \vec q\,, i(\vec\sigma_1\!+\!\vec\sigma_2)\!\cdot\!(\vec q\!\times\!\vec p\,)$ and $\vec\sigma_1\!\cdot\! \vec p\, \vec\sigma_2\!\cdot\!\vec p +\vec \sigma_1\!\cdot\! \vec p\,' \vec \sigma_2\!\cdot\!\vec p\,'$. For  notational convenience we use also the quadratic spin-orbit operator $\vec\sigma_1\!\cdot\!(\vec q\!\times\!\vec p\,)\vec\sigma_2\!\cdot\!(\vec q\!\times\!\vec p\,)$, (see eq.(2) in ref.\,\cite{vmedshort} for its decomposition). 

The present paper is organized as follows. In sections 2 and 3 we treat the $2\pi$-exchange and $2\pi1\pi$-exchange topologies. We recapitulate first the explicit expression for the respective 3N-interaction $V_\text{3N}$, which yet exhibits a kind of factorization in the momentum transfers $\vec q_{1,2,3}$ at the three nucleons. Then we present the results for $V_\text{med}$ as obtained by closing one nucleon-line and integrating the resulting in-medium loop: $-(2\pi )^{-3}\!\int\!d^3l\,\theta(k_f-|\vec l\,|)$. In this procedure we distinguish by superscripts those contributions to $V_\text{med}$, which originate from self-closings $(0)$, vertex corrections $(1), (2)$, and double exchanges $(3)$ (see figures 1-5 in ref.\cite{vmedshort} for the corresponding symbolic diagrams). Section 4 is devoted to the rather complicated ring topology. We start from the basic expression for $V_\text{3N}$ in the form of a three-dimensional loop-integral over pion-propagators and momentum-factors. In order to derive good semi-analytical expressions for the contributions to $V_\text{med}$, it essential to invert the order of the loop-integration and the integration over a Fermi sphere of radius $k_f$. The latter integral can be expressed in terms of elementary functions and for the former the angular part can still be solved analytically. This way we end up with a single-integral representation for each contribution to $V_\text{med}$ from the 3N-ring interaction. We present in three subsections the results from self-closings of nucleon-lines and from concatenations of two nucleon-lines for the parts of $V_\text{3N}$ proportional to $g_A^4$ and $g_A^6$. In the case of the isoscalar and isovector central potentials  a regularization by subtracting asymptotic constants is applied to get ultraviolet-convergent loop-integrals.  In section 5 we consider the $2\pi$-exchange 3N-interaction in the new notation of ref.\,\cite{twopi4}, where it is described by two structure functions $\tilde g_+(q_2)$ and $\tilde h_-(q_2)$ that are related to elastic $\pi N$-scattering amplitudes. In this form we derive general expressions for the contributions to $V_\text{med}$ as they arise from self-closings, pionic vertex corrections, and double exchanges. While these formulas are directly applicable to the longest-range 3N-interaction derived recently in chiral effective field theory with explicit $\Delta(1232)$-isobars \cite{twopidelta}, we perform in subsection 5.2 the calculation of $V_\text{med}$  from the relativistic $1/M$-correction to $V_\text{3N}$ obtained in this framework. Finally, subsection 5.3  discusses the matching of the new notation for the $2\pi$-exchange 3N-interaction to the original forms at N$^3$LO and N$^2$LO. The appendix contains a collection of relevant functions that were encountered in the course of the calculations.

In summary, after eventual partial-wave projection our results for $V_\text{med}$ are suitable for easy implementention of the subleading long-range chiral 3N-forces into nuclear many-body calculations. In the next step the calculation of $V_\text{med}$ should be extended to the intermediate-range components at N$^4$LO \cite{midrange4}. For the $2\pi1\pi$-exchange topology this amounts merely to a richer spin-momentum structure with twelve instead of eight terms, while for the ring diagrams proportional to $c_{1,2,3,4}$ one has to deal with genuine four-dimensional euclidean loop-integrals. Moreover, the input to nuclear many-body calculations should be completed at N$^3$LO by converting the chiral four-nucleon force \cite{4Nforce} into a density-dependent $V_\text{med}$. A fortunate property of the chiral 4N-interactions is that self-closings of nucleon-lines give zero in spin-saturated isospin-symmetric nuclear matter. But still a lot of calculational work is necessary to treat the large number of concatenations of nucleon-lines for all 4N-interaction terms, and to bring the results for $V_\text{med}$  into a good semi-analytical form.

\section{Two-pion exchange topology}
We start with the longest-range component of the subleading chiral 3N-interaction. It arises from $2\pi$-exchange with the corresponding symbolic diagram shown on the left of Fig.\,1. Using for simplification the relation $q_2^2 = q_1^2+q_3^2 +2\vec q_1 \!\cdot\!\vec q_3$, the expression in eq.(2.9) of ref.\,\cite{3Nlong} takes the form:
\begin{eqnarray}
2V_\text{3N}&=&{g_A^4 \over 128\pi f_\pi^6} {\vec\sigma_1\!\cdot\! \vec q_1 \vec
\sigma_3\!\cdot\! \vec q_3 \over (m_\pi^2+q_1^2) (m_\pi^2+q_3^2)}\Big\{\vec\tau_1
\!\cdot\! \vec\tau_3\big[m_\pi(m_\pi^2+q_1^2+q_3^2+2q_2^2) \nonumber \\ && +
(2m_\pi^2+q_2^2)(3m_\pi^2+q_1^2+q_3^2+2q_2^2)A(q_2) \big]   \nonumber \\ && +
\vec\tau_1\!\cdot\!(\vec\tau_2\!\times\! \vec\tau_3)\, \vec\sigma_2\!\cdot\!(
\vec q_1\!\times\!\vec q_3)\big[m_\pi+ (4m_\pi^2+q_2^2) A(q_2) \big]\Big\} \,, \end{eqnarray}
with the frequently occurring pion-loop function
\begin{equation}A(s) = {1\over 2s} \arctan{s\over 2m_\pi}\,.\end{equation}
Since it is symmetric under $1\!\leftrightarrow\!3$, we have multiplied here $V_\text{3N}$ with a factor $2$ to account already for this permutation.\footnote{In this pedantic notation the factor $2$ should also be written on the left hand side of eqs.(31,34) in ref.\cite{vmedshort}.} Alternatively, one could  distinguish the equivalent pion-couplings to nucleon $1$ and nucleon $3$ and the factor $2$ would emerge at the end of the calculation. The magnitudes of the momentum transfers $q_1, q_2, q_3$ satisfy the same inequalities as the three side-lengths of a triangle. Actually, $V_\text{3N}$ in eq.(1) comes from a single pion-loop diagram \cite{3Nlong} and the remaining corrections can be cast into shifts of the low-energy constants $c_{1,3,4}$ entering the leading-order chiral $2\pi$-exchange 3N-interaction:
\begin{equation} c_1\to c_1-{g_A^2m_\pi \over 64\pi f_\pi^2}\,, \qquad   c_3\to c_3+{g_A^4m_\pi \over 16\pi f_\pi^2}\,,  \qquad   c_4\to c_4-{g_A^4m_\pi \over 16\pi f_\pi^2}\,. \end{equation}  
\subsection{Contributions to in-medium NN-potential}
Now, we present the contributions of $V_\text{3N}$ in eq.(1) to the in-medium NN-potential $V_\text{med}$. The self-closing of nucleon-line $2$ gives (after relabeling $3\to 2$) the contribution
\begin{equation}V^{(0)}_\text{med} = - {g^4_A m_\pi k^3_f \over (4\pi f^2_\pi)^3}
{\vec\tau_1\!\cdot\!\vec\tau_2\over (m^2_\pi+q^2)^2} \,\vec\sigma_1\!\cdot\!\vec
q\, \vec\sigma_2\!\cdot\!\vec q\, \Big(q^2+{5m^2_\pi\over 6}\Big)\,. \end{equation}
From pionic vertex corrections on either nucleon-line one obtains the (total) contribution:
\begin{eqnarray}V^{(1)}_\text{med} &=&  {2g^4_A \over (8\pi f^2_\pi)^3}
{\vec\tau_1\!\cdot\!\vec\tau_2\over m^2_\pi+q^2} \,\vec\sigma_1\!\cdot\!\vec
q\, \vec\sigma_2\!\cdot\!\vec q\,\Big\{m_\pi\big[ 2k_f^3 -8\Gamma_2-2m_\pi^2
(\Gamma_0+\Gamma_1) \nonumber \\  &&  +q^2 (\Gamma_0-\Gamma_1-2\Gamma_3)\big]  +
S_0(p) + J_1(p,q)+ J_2(p,q)\Big\}\,,\end{eqnarray}
where the functions $\Gamma_\nu(p,k_f)$ are defined in the appendix of
ref.\,\cite{vmedshort}. The part linear in $m_\pi$ comes obviously from the first line in eq.(1) and the decomposition $S_0(p) + J_1(p,q)+ J_2(p,q)$ is obtained by canceling momentum factors against a pion-propagator. The three new functions appearing at the end of eq.(5) read:
\begin{eqnarray}S_0(p)  &=& \bigg[ {p^4\over 105}+{2 p^2 \over 15}(
4m_\pi^2 -k_f^2) -{k_f^3 \over 3}(p+k_f)- 2 k_f^2 m_\pi^2 - 4 m_\pi^4\nonumber \\
 &&-{k_f^3 \over 3p}\Big({2k_f^2\over 5}+5m_\pi^2\Big)+{k_f^5 \over 5 p^3}\Big({k_f^2\over 21}+m_\pi^2\Big)\bigg] \arctan{p+ k_f \over 2m_\pi} \nonumber \\ && +\bigg[ {2 p^2 \over 15}( k_f^2-4m_\pi^2)-{p^4\over 105}+ {k_f^3 \over 3}(k_f-p)+ 2 k_f^2 m_\pi^2 +4 m_\pi^4 \nonumber \\ &&-{k_f^3 \over 3p}\Big({2k_f^2\over 5} +5m_\pi^2\Big)+{k_f^5 \over 5 p^3}\Big({k_f^2\over 21}+m_\pi^2\Big)\bigg] \arctan{p- k_f \over 2m_\pi}  \nonumber \\ && +m_\pi^3 \bigg[
{17\over p}\bigg({m_\pi^2 \over 15}+ {k_f^2 \over 12}\bigg) -{29 p\over 24} -{1
\over p^3}\bigg( {5k_f^4 \over 24}+  {k_f^2 m_\pi^2 \over 5}+ {2m_\pi^4 \over 35}
\bigg) \bigg] \ln{4m_\pi^2+(p+k_f)^2\over 4m_\pi^2+(p-k_f)^2}\nonumber \\ && +
{k_f m_\pi \over 210}\bigg( 599 m_\pi^2-316k_f^2  -8 p^2+{12 m_\pi^4+39k_f^2m_\pi^2 -8 k_f^4\over p^2}\bigg)\,, \end{eqnarray}
\begin{eqnarray}J_1(p,q) &=& {1\over 4q^2} \int_{p-k_f}^{p-k_f}\!ds
(2m_\pi^2+s^2)(2m_\pi^2+q^2 +2s^2) \arctan{ s\over 2m_\pi}\nonumber \\ && \times
\bigg\{{k_f^2 -(p-s)^2 \over p} + {q^2-m_\pi^2-s^2 \over q} \ln{ q X + 2\sqrt{W}
\over (2p+q)[m_\pi^2+(q-s)^2]}\bigg\} \,,  \end{eqnarray}
\begin{eqnarray}J_2(p,q) &=& {1\over 4q^2}\int_{p-k_f}^{p-k_f}\!ds
\Big[2m_\pi s+(4m_\pi^2+s^2) \arctan{ s\over 2m_\pi} \Big] \bigg\{ {1\over p}(q^2-m_\pi^2-s^2)\big[k_f^2 -(p-s)^2 \big]\nonumber \\ && +{1\over q}\big[m_\pi^2+(q+s)^2\big]\big[m_\pi^2+(q-s)^2\big]\ln{ q X + 2\sqrt{W} \over (2p+q)[m_\pi^2+(q-s)^2]}   \bigg\}\,,  \end{eqnarray}
with the auxiliary polynomials
\begin{eqnarray} && X = m_\pi^2 +2(k_f^2-p^2)+q^2-s^2\,, \nonumber \\
&& W = k_f^2 q^4+p^2(m_\pi^2+s^2)^2 +q^2\big[ (k_f^2-p^2)^2+m_\pi^2(k_f^2+p^2)
   -s^2(k_f^2+p^2+m_\pi^2)\big]\,.  \end{eqnarray}
The above representations of $J_1(p,q)$ and $J_2(p,q)$ arise from the reduction of Fermi sphere integrals  $(2\pi)^{-1}\!\!\int_{|\vec l\,|<k_f} \!d^3 l \,F(|\vec l + \vec p\,|)[m_\pi^2+ (\vec l +  \vec p\,')^2]^{-1}\{1,\vec l\,\}$ over even functions $F(s)= F(-s)$ to one-dimensional integrals $\int_{p-k_f}^{p+k_f}\!ds ...F(s)$. Finally, the two diagrams related to double exchange lead to the expression:
 \begin{eqnarray} V^{(3)}_\text{med} & = &{g^4_A\over (8\pi  f^2_\pi)^3} \bigg\{
\big[ m_\pi+(2m_\pi^2+q^2) A(q) \big]\big[8k_f^3+6q^2(\Gamma_0-\Gamma_1)
\big] -6m_\pi^2\Gamma_0\big[3m_\pi\nonumber\\ && + (2m_\pi^2+q^2) A(q)\big]+
3(2m_\pi^2+q^2)\big[m_\pi^3-2m_\pi q^2-(2m_\pi^4+5m_\pi^2q^2+2q^4)A(q)\big] G_{0}
 \nonumber\\ && + 3i( \vec\sigma_1\!+\!\vec\sigma_2)\!\cdot\!(\vec q\!\times \!
 \vec p\,)\Big\{ 2\big[m_\pi+(2m_\pi^2+q^2)A(q) \big](\Gamma_0+ \Gamma_1)\nonumber\\ &&+\big[2m_\pi q^2-m_\pi^3+(2m_\pi^4+5m_\pi^2q^2+2q^4)A(q)\big](G_0+2G_1)
\Big\}\nonumber\\ && +\vec\tau_1\!\cdot\!\vec\tau_2 \big[m_\pi+(
4m_\pi^2+q^2)A(q)\big]\Big\{ i(\vec\sigma_1\!+\!\vec\sigma_2)\!\cdot\!(\vec q\!
\times\! \vec p\,)\big[2\Gamma_0+2\Gamma_1 -(2m_\pi^2+q^2)(G_0+2G_1)\big]
\nonumber\\ && +4(\vec\sigma_1\!\cdot\!\vec q\, \vec\sigma_2\!\cdot\!\vec q-
\vec\sigma_1\!\cdot\!\vec \sigma_2 q^2) G_2- 4\vec\sigma_1\!\cdot\!(\vec q\!
\times\! \vec p\,)\vec \sigma_2\!\cdot\!(\vec q\!\times\!\vec p\,)\big[G_0+
4G_1 +4G_3\big]\Big\} \bigg\} \,,  \end{eqnarray}
with the functions $G_\nu(p,q,k_f)$ defined in the appendix (for alternative relations, see also the appendix of ref.\cite{vmedshort}). 
\section{Two-pion-one-pion exchange topology}
Next, we come to the $2\pi1\pi$-exchange three-nucleon interaction represented by the middle diagram in Fig.\,1. According to eqs.(2.16)-(2.20) in ref.\cite{3Nlong} this chiral 3N-interaction  can be written in the form:
\begin{eqnarray}
V_\text{3N}&=&{g_A^4 \over 256\pi f_\pi^6} { \vec\sigma_3\!\cdot\! \vec q_3 \over m_\pi^2+q_3^2}\Big\{ \vec\tau_1 \!\cdot\! \vec\tau_3\big[\vec\sigma_2\!\cdot\! \vec q_1\,\vec q_1\!\cdot\! \vec q_3\, f_1(q_1) + \vec\sigma_2\!\cdot\! \vec q_1\,f_2(q_1)+ \vec\sigma_2\!\cdot\! \vec q_3\,f_3(q_1)\big] \nonumber\\ && +
\vec\tau_2\!\cdot\! \vec\tau_3\big[\vec\sigma_1\!\cdot\! \vec q_1\,\vec q_1\!\cdot\! \vec q_3\, f_4(q_1) + \vec\sigma_1\!\cdot\! \vec q_3\,f_5(q_1)+ \vec\sigma_2\!\cdot\! \vec q_1\,f_6(q_1)+ \vec\sigma_2\!\cdot\! \vec q_3\,f_7(q_1)\big] \nonumber\\ && +(\vec\tau_1\!\times \!\vec\tau_2)\!\cdot\! \vec\tau_3\, (\vec\sigma_1\!\times\!\vec\sigma_2)\!\cdot\!\vec q_1\,f_8(q_1)\Big\}\,,  \end{eqnarray}
where we have pulled out a common factor $g_A^4/(256\pi f_\pi^6)$, and thus the reduced  functions $f_j(s)$ read: 
\begin{eqnarray}&& f_1(s) = {m_\pi\over s^2}(1-2 g_A^2)-{g_A^2m_\pi\over 4m_\pi^2 + s^2}  +\Big[1+g_A^2 +{4 m_\pi^2 \over s^2}(2 g_A^2-1)\Big] A(s) \,,
\\ && f_2(s)= f_7(s)={1\over 2} f_6(s)= 2m_\pi+(4m_\pi^2+2s^2)A(s) \,,
\\ && f_3(s)= m_\pi(1-3g_A^2)+\big[4m_\pi^2(1-2g_A^2)+s^2(1-3g_A^2)\big]A(s) \,, \\  &&  f_5(s)=-s^2 f_4(s)=2 g_A^2 s^2 A(s)\,, \\ && f_8(s)= -{1\over 2}\big[m_\pi+(4m_\pi^2+s^2)A(s)\big] \,, \end{eqnarray}
with $A(s)$ defined in eq.(2). In perspective we note that at sub-subleading order \cite{midrange4} the chiral $2\pi1\pi$-exchange 3N-interaction has a richer spin- and momentum-dependence and twelve functions $f_j(q_1)$ are needed to represent all diagrams belonging to this topology at N$^4$LO.
\subsection{Contributions to in-medium NN-potential}
Again, we list the four contributions from $V_\text{3N}$ in eq.(11) to the in-medium NN-potential $V_\text{med}$. Only the self-closing of nucleon line 1 gives a nonvanishing spin-isospin trace, and with $f_7(0) =3m_\pi $ one obtains (after relabeling $3\to 1$) the contribution: 
\begin{equation}V^{(0)}_\text{med} = {g^4_A m_\pi k^3_f \over (4\pi f^2_\pi)^3}
{\vec\tau_1\!\cdot\!\vec\tau_2\over m^2_\pi+q^2} \,\vec\sigma_1\!\cdot\!\vec
q\, \vec\sigma_2\!\cdot\!\vec q\,, \end{equation}
which is of the form: one-pion exchange NN-interaction times a factor linear in density $\rho=2k_f^3 /3\pi^2$. On the other hand the vertex corrections by $1\pi$-exchange, incorporated in eq.(11) through the second factor $\vec\sigma_3\!\cdot\! \vec q_3/(m_\pi^2+q_3^2)$, produce the contribution:
\begin{eqnarray}V^{(1)}_\text{med} &=&  {g_A^4 \over (8\pi f_\pi^2)^3} \bigg\{ 
\Big(2m_\pi^2\Gamma_0-{4k_f^3\over 3}\Big)\big[\vec\tau_1\!\cdot\!\vec\tau_2 f_3(q)+3f_7(q) \big] - \vec\tau_1\!\cdot\!\vec\tau_2 \Big(2\Gamma_2 +{q^2 \over 2} \widetilde\Gamma_3\Big)  q^2 f_1(q)
\nonumber\\ && + \widetilde\Gamma_1 q^2 \big[3 f_6(q) + \vec \tau_1\!\cdot\!\vec\tau_2 f_2(q)\big]-2\vec\sigma_1\!\cdot\!\vec\sigma_2\big[3 \Gamma_2 f_5(q)+
\vec\tau_1\!\cdot\!\vec\tau_2 \widetilde\Gamma_1 q^2 f_8(q) \big] \nonumber\\ && 
-3 (\vec\sigma_1\!\cdot\!\vec p\, \vec\sigma_2\!\cdot\!\vec p+\vec\sigma_1\!\cdot\!\vec p\,'\, \vec\sigma_2\!\cdot\!\vec p\,')\widetilde \Gamma_3 f_5(q) + 2\vec\sigma_1\!\cdot\!\vec q\, \vec\sigma_2\!\cdot\!\vec q\, \Big[\vec\tau_1\!\cdot\!\vec\tau_2 \widetilde\Gamma_1 f_8(q)-3\Big(\Gamma_2 +{q^2 \over 4} \widetilde\Gamma_3\Big) f_4(q) \Big] \nonumber\\ && +i( \vec\sigma_1\!+\!\vec\sigma_2)\!\cdot \! (\vec q\!\times \! \vec p\,) \Big[\vec\tau_1\!\cdot\!\vec\tau_2 \Big(\widetilde\Gamma_1\big[2f_8(q)-f_2(q)\big] +\widetilde\Gamma_3{q^2 \over 2} f_1(q)\Big)-3 \widetilde\Gamma_1 f_6(q) \Big] \bigg\}\,. \end{eqnarray}
Here, we have introduced the frequently occuring combinations $\widetilde\Gamma_1(p)= \Gamma_0(p)+\Gamma_1(p)$ and $\widetilde\Gamma_3(p)= \Gamma_0(p) +2\Gamma_1(p)+\Gamma_3(p)$. Moreover, the vertex corrections by $2\pi$-exchange (compiled in the expression in curly brackets of eq.(11)) can be summarized as the one-pion exchange NN-interaction times a $(p,q,k_f)$-dependent factor 
\begin{equation}V^{(2)}_\text{med} = {g^4_A \over (8\pi f^2_\pi)^3}
{\vec\tau_1\!\cdot\!\vec\tau_2\over m^2_\pi+q^2} \,\vec\sigma_1\!\cdot\!\vec
q\, \vec\sigma_2\!\cdot\!\vec q\,\Big\{ S_1(p) +g_A^2 S_2(p)+q^2\big[S_3(p) +g_A^2 S_4(p)\big]\Big\}\,, \end{equation}
with the four functions $S_{1,2,3,4}(p,k_f)$ given by:
\begin{eqnarray}
S_1(p)&\!\!\!\!=\!\!\!\!&\bigg[{k_f^3 p\over 3}-{k_f^5\over 5 p^3}\Big({k_f^2\over 7} +2m_\pi^2\Big)+{2k_f^3\over 3p}\Big({2k_f^2\over 5}+m_\pi^2\Big)+{k_f^4\over 2}+{p^2\over 15} (k_f^2+4m_\pi^2)+{p^4\over 210} \bigg]\arctan{p+k_f \over 2m_\pi}\nonumber\\ && +\bigg[{k_f^3p\over 3}-{k_f^5\over 5 p^3}\Big({k_f^2\over 7} +2m_\pi^2\Big)+{2k_f^3\over 3p}\Big({2k_f^2\over 5}+m_\pi^2\Big)-{k_f^4\over 2}-{p^2\over 15} (k_f^2+4m_\pi^2)-{p^4\over 210} \bigg]\arctan{p-k_f \over 2m_\pi}\nonumber\\ && + {m_\pi^3 \over p^3} \bigg[{k_f^4\over 4} - {2m_\pi^2\over 5}\Big(k_f^2 + {18m_\pi^2\over 7}\Big) + {p^2\over 3} \Big({k_f^2\over 2} - {14 m_\pi^2 \over 5}\Big)- {5 p^4\over 12}\bigg] \ln{4m_\pi^2+(p+k_f )^2 \over 4m_\pi^2+(p-k_f)^2}\nonumber\\ && + {k_f m_\pi\over 35} \bigg[{110k_f^2+71m_\pi^2-2p^2 \over 3}+ {4 k_f^4 + 5 k_f^2 m_\pi^2 + 36 m_\pi^4\over p^2}\bigg]\,,  \end{eqnarray}
\begin{eqnarray}
S_2(p)&\!\!\!\!=\!\!\!\!&{1\over 3}\bigg[k_f^3 p-{k_f^5\over 5 p^3}\Big({k_f^2\over 7} +8m_\pi^2\Big)+{2k_f^3\over p}\Big({k_f^2\over 5}+4 m_\pi^2 \Big)+k_f^4\nonumber\\ && +8m_\pi^2(k_f^2+4m_\pi^2)+{2p^2\over 5} (k_f^2-4m_\pi^2)-{p^4\over 35} \bigg]\arctan{p+k_f \over 2m_\pi}\nonumber\\ && +{1\over 3}\bigg[k_f^3 p-{k_f^5\over 5 p^3}\Big({k_f^2\over 7} +8m_\pi^2\Big)+{2k_f^3\over p}\Big({k_f^2\over 5}+4 m_\pi^2 \Big)-k_f^4\nonumber\\ && -8m_\pi^2(k_f^2+4m_\pi^2)+{2p^2\over 5} (4m_\pi^2-k_f^2)+{p^4\over 35} \bigg]\arctan{p-k_f \over 2m_\pi} \nonumber\\ && + {m_\pi \over 3} \bigg[{p\over 8}(p^2-3k_f^2)+7m_\pi^2 p+{1\over p}\Big({3k_f^4\over 8} -8k_f^2m_\pi^2-{62m_\pi^4\over 5}\Big)\nonumber\\ && + {1\over p^3}\Big(k_f^4m_\pi^2-{k_f^6\over 8}-{2k_f^2m_\pi^4\over 5}-{64m_\pi^6\over 35} \Big)\bigg] \ln{4m_\pi^2+(p+k_f )^2 \over 4m_\pi^2+(p-k_f)^2}\nonumber\\ && +{k_f m_\pi\over 105} \bigg[{1244\over 3} k_f^2- 702m_\pi^2 + {43 k_f^4 \over 2p^2} + {2m_\pi^2\over p^2}(32m_\pi^2-k_f^2)- {27 p^2\over 2}\bigg]\,,  \end{eqnarray}
\begin{eqnarray}S_3(p)&\!\!\!\!=\!\!\!\!&\bigg[{2m_\pi^2-k_f^2 \over 15}+{p^2\over 105} -{k_f^3\over 12p}+{k_f^3\over 3p^3}\Big({k_f^2\over 10}+m_\pi^2\Big)-{k_f^5\over 5p^5}\Big({k_f^2\over 28}+m_\pi^2\Big) \bigg]\arctan{p+k_f \over 2m_\pi}\nonumber \\ && + \bigg[{k_f^2-2m_\pi^2\over 15}-{p^2\over 105} -{k_f^3\over 12p}+{k_f^3\over 3p^3}\Big({k_f^2\over 10}+m_\pi^2\Big)-{k_f^5\over 5p^5}\Big({k_f^2\over 28}+m_\pi^2\Big) \bigg]\arctan{p-k_f \over 2m_\pi}
\nonumber\\ && + {m_\pi^3 \over p^5} \bigg[{k_f^4\over 4} + {2m_\pi^2\over 5}\Big(k_f^2 + {6m_\pi^2\over 7}\Big) + {p^2\over 3} \Big( {2m_\pi^2 \over 5} -{k_f^2\over 2}\Big)- { p^4\over 12}\bigg] \ln{4m_\pi^2+(p+k_f )^2 \over 4m_\pi^2+(p-k_f)^2}\nonumber\\ && + {k_f m_\pi\over 35p^2} \bigg[{k_f^4-11k_f^2m_\pi^2-12m_\pi^4 \over p^2}- {11 k_f^2 + 5 m_\pi^2 + 4p^2\over 3}\bigg]\,,  \end{eqnarray}
\begin{eqnarray}S_4(p)&\!\!\!\!=\!\!\!\!&{1\over 3}\bigg[{k_f^2-4m_\pi^2 \over 5}-{p^2\over 35} +{k_f^3\over 4p}-{k_f^3\over p^3}\Big({k_f^2\over 10}+2m_\pi^2\Big)+{3k_f^5\over 5p^5}\Big({k_f^2\over 28}+2m_\pi^2\Big) \bigg]\arctan{p+k_f \over 2m_\pi}\nonumber \\ && + {1\over 3}\bigg[{4m_\pi^2-k_f^2\over 5}+{p^2\over 35} +{k_f^3\over 4p}-{k_f^3\over p^3}\Big({k_f^2\over 10}+2m_\pi^2\Big)+{3k_f^5\over 5p^5}\Big({k_f^2\over 28}+2m_\pi^2\Big) \bigg]\arctan{p-k_f \over 2m_\pi}\nonumber \\ && + {m_\pi \over 4} \bigg[{k_f^4\over p^5}\Big({k_f^2\over 8}-m_\pi^2\Big)+{2m_\pi^4 \over 5p^5}\Big(k_f^2+{32m_\pi^2\over 7}\Big)-{3k_f^4\over 8 p^3} +{2m_\pi^2 \over 3p^3}\Big(k_f^2+{m_\pi^2\over 5}\Big)+{3k_f^2\over 8p} +{m_\pi^2\over 3p}-{p\over 8} \bigg] \nonumber\\ && \times \ln{4m_\pi^2+(p+k_f )^2 \over 4m_\pi^2+(p-k_f)^2}+ {k_f m_\pi\over 35} \bigg[{137\over 24} +{1\over p^4}\Big({k_f^2m_\pi^2\over 2}-16m_\pi^4-{43k_f^4 \over 8}\Big) + {92 k_f^2+17m_\pi^2 \over 6p^2}\bigg]\,.  \end{eqnarray}
Finally, the more complicated contribution from double exchange (in fact it is $3\pi$-exchange) reads:
\begin{eqnarray}V^{(3)}_\text{med}&\!\!=\!\!&{g_A^4\over (8\pi f_\pi^2)^3}\bigg\{ 
3\bigg[\vec\sigma_1\!\cdot\!\vec \sigma_2\Big(2I_{2,2}-2I_{3,2}-H_{1,2}-\tilde I_{1,2}\Big)+ \vec\sigma_1\!\cdot\!\vec q\, \vec \sigma_2\!\cdot\!\vec q \,\Big({H_{1,1}+\tilde I_{1,4}\over 2}-I_{2,4}-I_{3,5}\Big)\nonumber\\ && +(\vec\sigma_1\!\cdot\!\vec p\, \vec\sigma_2\!\cdot\!\vec p+\vec\sigma_1\!\cdot\!\vec p\,'\, \vec\sigma_2\!\cdot\!\vec p\,')\Big(I_{2,3}-I_{3,3}-{H_{1,3}+\tilde I_{1,3}\over 2}\Big)\bigg]+ \vec\tau_1\!\cdot\!\vec \tau_2\bigg[2m_\pi^2 I_{5,0}-2H_{5,0}\nonumber\\ && +{q^2 \over 2}(H_{4,1}\!+\!\tilde I_{4,4})-p^2(H_{4,3}\!+\!\tilde I_{4,3})-3H_{4,2}-3 \tilde I_{4,2} -i(\vec\sigma_1\!+\!\vec\sigma_2)\!\cdot\!(\vec p\!\times\!\vec q\,)\Big({H_{4,1}\!+\!\tilde I_{4,1}\over 2}+2I_{8,1}\Big) \nonumber\\ && +2\vec \sigma_1\!\cdot\!\vec \sigma_2\Big(I_{6,2}-I_{7,2}-4 I_{8,2}-2p^2I_{8,3}+q^2I_{8,4}\Big) 
-\vec\sigma_1\!\cdot\!\vec q\, \vec \sigma_2\!\cdot\!\vec q \,\Big(I_{6,4}+ I_{7,5}+2I_{8,4}\Big) \nonumber\\ && +(\vec\sigma_1\!\cdot\!\vec p\, \vec\sigma_2\!\cdot\!\vec p+\vec\sigma_1\!\cdot\!\vec p\,'\, \vec\sigma_2\!\cdot\!\vec p\,')\Big(I_{6,3}-I_{7,3}+2I_{8,3}\Big)\bigg]\bigg\}\,.  \end{eqnarray}
The double-indexed functions $H_{j,\nu}(p)$ for $j=1,4,5$ are defined by:
\begin{equation} H_{j,0}(p) = {1\over 2p}\int_{p-k_f}^{p+k_f}\!\!ds\, s f_j(s)\big[k_f^2-(p-s)^2\big]\,, \end{equation} 
\begin{equation} H_{j,1}(p) = {1\over 8p^3}\int_{p-k_f}^{p+k_f}\!\!ds\, s f_j(s)\big[k_f^2-(p-s)^2\big]\big[(p+s)^2-k_f^2\big]\,, 
\end{equation} 
\begin{equation} H_{,j2}(p) = {1\over 48p^3}\int_{p-k_f}^{p+k_f}\!\!ds\, s f_j(s)\big[k_f^2-(p-s)^2\big]^2(s^2+4s p +p^2-k_f^2)\,, 
\end{equation} 
\begin{equation} H_{j,3}(p) = {1\over 16p^5}\int_{p-k_f}^{p+k_f}\!\!ds\, s f_j(s)\big[k_f^2-(p-s)^2\big]\big[(p+s)^2-k_f^2\big](p^2+s^2-k_f^2)\,, \end{equation} 
and could still be solved analytically in terms of $\arctan[(p\pm k_f)/2m_\pi]$ and $\ln[4m_\pi^2 +(p\pm k_f)^2]$. Two examples are given in the appendix.
The other double-indexed functions  $I_{j,\nu}(p,q)$  are defined by: 
\begin{equation} I_{j,0}(p,q) = {1\over 2q}\int_{p-k_f}^{p+k_f}\!\!ds\, s f_j(s)
\ln{q X +2\sqrt{W} \over (2p+q)[m_\pi^2+(s-q)^2]}\,, \end{equation} \begin{equation} I_{j,1}(p,q) = {1\over 4p^2-q^2}\int_{p-k_f}^{p+k_f}\!\!ds\, s f_j(s)\bigg[ {p(s^2+m_\pi^2)-\sqrt{W} \over q^2}+{p^2+k_f^2-s^2 \over 2p}\bigg]\,, \end{equation} 
\begin{eqnarray} I_{j,2}(p,q) &\!\!\!\!\!\!\!\!=\!\!\!\!\!\!\!\!& {1\over 8q^2}\int_{p-k_f}^{p+k_f}\!\!ds\, s f_j(s)\bigg\{s(m_\pi^2+s^2+q^2)-p\Big(m_\pi^2+s^2+{3q^2 \over 4}\Big)\nonumber\\ && -{1\over 2q}\big[m_\pi^2+(s+q)^2\big] \big[m_\pi^2+(s-q)^2\big] \ln{q X +2\sqrt{W} \over (2p+q)[m_\pi^2+(s-q)^2]}  \nonumber\\ && - {X \sqrt{W} \over 4p^2-q^2} -{(k_f^2-s^2)^2 \over p} +{p\over  4p^2-q^2}\Big(m_\pi^2+2k_f^2 -s^2+{q^2\over 2}\Big)^2  \bigg\}\,, \end{eqnarray}
\begin{eqnarray} I_{j,3}(p,q) &\!\!\!\!\!\!\!\!=\!\!\!\!\!\!\!\!& {1\over (4p^2-q^2)^2}\int_{p-k_f}^{p+k_f}\!\!ds\, s f_j(s)\bigg\{ {X \sqrt{W} \over q^2} +{q^2 \over 8p^3}(k_f^2-s^2)^2 -{3pq^2\over 8}\nonumber\\ &&+{p \over q^2}(s^2+m_\pi^2)(2p^2+s^2-2k_f^2-m_\pi^2) +{p \over 2}(2k_f^2-3m_\pi^2+p^2-s^2) \nonumber\\ && +{1\over 4p}\Big[s^2(2m_\pi^2+q^2-4s^2)+k_f^2(10s^2-2m_\pi^2-3q^2)-6k_f^4\Big]  \bigg\}\,, \end{eqnarray}
\begin{eqnarray} I_{j,4}(p,q) &\!\!\!\!\!\!\!\!=\!\!\!\!\!\!\!\!& {1\over 4q^4}\int_{p-k_f}^{p+k_f}\!\!ds\, s f_j(s)\bigg\{ {X\over (4p^2-q^2)^2}\Big[\sqrt{W}(3q^2-4p^2)- 8p^3X\Big]\nonumber\\ && + \bigg[ {q^3\over 2}+q(s^2-m_\pi^2) -{3\over 2q}(s^2+m_\pi^2)^2\bigg] \ln{q X +2\sqrt{W} \over (2p+q)[m_\pi^2+(s-q)^2]} \nonumber\\ && +{p \over 4p^2-q^2}\Big[16k_f^4+8k_f^2(2m_\pi^2+q^2-2s^2 )+3m_\pi^4+3s^4+2q^2(m_\pi^2-s^2)-10s^2m_\pi^2\Big] \nonumber\\ && +s(3s^2-q^2+3m_\pi^2)+2p^3-p(4k_f^2+4m_\pi^2+q^2) +{2\over p}(k_f^2+q^2-s^2)(s^2-k_f^2)  \bigg\}\,,  \end{eqnarray}
\begin{equation} I_{j,5}(p,q) =-I_{j,4}(p,q)+{1\over 2q^2}\int_{p-k_f}^{p+k_f}\!\!ds\, s f_j(s)\bigg[{q^2\!-\!s^2\!-\!m_\pi^2\over q} \ln{q X +2\sqrt{W} \over (2p+q)[m_\pi^2+(s-q)^2]}+{k_f^2-(p-s)^2\over p}\bigg]\,. \end{equation} 
Furthermore, the functions  $\tilde I_{j,\nu}(p,q)$ with $j = 1,4$ appearing in eq.(24) are computed analogously by substituting in the integrand $f_j(s)$ by  $\tilde f_j(s)= (s^2-m_\pi^2-q^2)f_j(s) $.  Let us also explain the meaning of our nomenclature. The (second) index $\nu = 0$ labels scalar integrals, $\nu = 2$ the $\delta_{ij}$-part of $l_il_j$-tensor integrals, and $\nu= 1,3,4,5$ refer to conveniently chosen linear combinations of functions that arise from scalar, vector and tensor integrals, applying suitable projection techniques. We note as an aside that at $q=0$ one gets the relation $I_{j,\nu}(p,0) = \bar H_{j,\nu}(p)$ for $\nu = 0,1,2,3$, where $\bar H_{j,\nu}(p)$ is calculated according to eqs.(25-28) with $\bar f_j(s) =  f_j(s)/(m_\pi^2+s^2)$. Returning to the result for $V_\text{med}^{(3)}$ in eq.(24), it should be stated that the decomposition into $H_{j,\nu}$ and  $I_{j,\nu}$ is obtained by canceling momentum-factors against a pion-propagator, while  $\tilde I_{j,\nu}$ takes care of $s^2$-dependent remainder terms. Altogether, we have achieved a well manageable representation of $V_\text{med}$ in terms of functions which require at most one numerical integration.   

\section{Ring topology}
The three-nucleon ring interaction, represented by the right diagram in Fig.\,1, is generated by a circulating pion that gets absorbed and reemitted at each of the three nucleons. It possesses a rather complicated structure, because any factorization property in the three momentum transfers $\vec q_{1,2,3}$ is lost. We start with the basic expression for $ V_\text{3N}$ in the form of a three-dimensional loop-integral over pion-propagators and momentum-factors \cite{3Nlong}:
\begin{eqnarray}V_\text{3N}&\!\!\!\!=\!\!\!\!&{g_A^4\over 32f_\pi^6}\int\!{d^3l_2\over (2\pi)^3} {1\over (m_\pi^2+l_1^2)(m_\pi^2+l_2^2)(m_\pi^2+l_3^2)} \bigg\{2\vec \tau_1\!\cdot\!\vec \tau_2\Big[\vec l_1\!\cdot\!\vec l_2\, \vec l_2\!\cdot\!\vec l_3 -\vec\sigma_1\!\cdot\!(\vec l_2\!\times\!\vec l_3) \vec\sigma_3\!\cdot\!(\vec l_1\!\times\!\vec l_2)\Big]\nonumber\\ &&+\vec\tau_1\!\cdot\!(\vec \tau_2\!\times\!\vec \tau_3) \vec\sigma_1\!\cdot\!(\vec l_2\!\times\!\vec l_3)\, \vec l_1\!\cdot\!\vec l_2 +{g_A^2 \over m_\pi^2+l_2^2} \Big[-4\vec \tau_1\!\cdot\!\vec \tau_2\, \vec\sigma_2\!\cdot\!(\vec l_1\!\times\!\vec l_3) \vec\sigma_3\!\cdot\!(\vec l_1\!\times\!\vec l_2)\vec l_2\!\cdot\!\vec l_3 \\ && -2\vec \tau_1\!\cdot\!\vec \tau_3\, \vec l_1\!\cdot\!\vec l_2\, \vec l_1\!\cdot\!\vec l_3\,\vec l_2\!\cdot\!\vec l_3+ \vec\tau_1\!\cdot\!(\vec \tau_2\!\times\!\vec \tau_3) \vec\sigma_2\!\cdot\!(\vec l_1\!\times\!\vec l_3)\vec l_1\!\cdot\!\vec l_2\,\vec l_2\!\cdot\!\vec l_3 +3 \vec\sigma_1\!\cdot\!(\vec l_2\!\times\!\vec l_3)\vec\sigma_3\!\cdot\!(\vec l_1\!\times\!\vec l_2)\vec l_1\!\cdot\!\vec l_3\Big]\bigg\}\,,\nonumber \end{eqnarray}
where one has to set $\vec l_1= \vec l_2-\vec q_3$ and $\vec l_3= \vec l_2+\vec q_1$. The $g_A^6$-part is explicitly written in eqs.(2.25,2.26) of ref.\,\cite{3Nlong} and the $g_A^4$-part is easily deduced from the corresponding coordinate-space potential in eq.(2.29). In fact the 3N-ring interaction $V_\text{3N}$ has been evaluated and expressed in the appendix of ref.\,\cite{3Nlong} in terms of seven $S$-functions and eleven $R$-functions,\footnote{As noticed in ref.\,\cite{midrange4}, additional symmetry factors $1/2$ were missing for $R_{6,8,9,10,11}$ in eq.(A1) of ref.\,\cite{3Nlong}.} which are each composed of $A(q_{1,2,3})$ and the 3-dimensional euclidean three-point function: 
\begin{eqnarray}J(q_1,q_2,q_3) &= & \int\!{d^3l_2\over (2\pi)^3} {1\over [m_\pi^2+(\vec l_2-\vec q_3)^2](m_\pi^2+\vec l_2^{\,2})[m_\pi^2+(\vec l_2+\vec q_1)^2]}
\nonumber\\ 
&=&{1\over 4\pi} \int_{2m_\pi}^\infty \!\! d\mu\, {\mu \over \mu^2+q_2^2}\Big\{ (q_1q_3 \mu)^2+m_\pi^2\big[\mu^2+(q_1+q_3)^2\big] \big[\mu^2+(q_1-q_3)^2\big]\Big\}^{-1/2} \,,
\end{eqnarray}
where the second line gives its dispersive representation, and $2\vec q_1\!\cdot\!\vec q_3 = q_2^2-q_1^2-q_3^2$ has been used. The latter dispersion-integral can actually be solved analytically in the form:
\begin{eqnarray} J(q_1,q_2,q_3) &=& {1\over 4\pi \sqrt{\Sigma} } \arctan{ \sqrt{\Sigma} \over 8m_\pi^3+(q_1^2+q_2^2+q_3^2)m_\pi}\nonumber\\  &=&  {1\over 4\pi \sqrt{\Sigma} } \arccos{8m_\pi^3+(q_1^2+q_2^2+q_3^2)m_\pi\over \sqrt{(4m_\pi^2+q_1^2) (4m_\pi^2+q_2^2)(4m_\pi^2+q_3^2)}}\,,
\end{eqnarray}
with the polynomial $\Sigma = (q_1 q_2q_3)^2+m_\pi^2(q_1+q_2+q_3)(q_1+q_2-q_3)(q_1+q_3-q_2)(q_2+q_3-q_1)>0$. The analogous loop-function $\widetilde J(q_1,q_2,q_3)$, with the middle pion-propagator in eq.(36) squared, reads:
\begin{eqnarray}\widetilde J(q_1,q_2,q_3) &=& {1\over 4\pi \Sigma}\bigg\{{q_2^2\over 2\sqrt{\Sigma} }(q_1^2+q_3^2-q_2^2)  \arctan{ \sqrt{\Sigma} \over 8m_\pi^3+(q_1^2+q_2^2+q_3^2)m_\pi}\nonumber\\  && +(4m_\pi^2+q_2^2)\bigg[ {1\over 2m_\pi}-{m_\pi \over 4m_\pi^2+q_1^2}-{m_\pi \over 4m_\pi^2+q_3^2}\bigg] -{m_\pi q_3^2 \over 4m_\pi^2+q_1^2}-{m_\pi q_1^2 \over 4m_\pi^2+q_3^2}\bigg\}\,. \end{eqnarray}
In order compute the contributions to the in-medium potential $V_\text{med}$ from the 3N-ring interaction $V_\text{3N}$ in eq.(35) in a useful semi-analytical form, it turns out to be crucial to invert the order of the loop-integration $\int\!d^3l_2$ and the integration over a Fermi sphere of radius $k_f$. After this rearrangement the Fermi sphere integral goes over a pion-propagator (or its square) and can be solved by the $\Gamma_\nu(\gamma_\nu)$-functions \cite{vmedshort}. At the same time the loop-momentum can be freely shifted, since it is an unconstrained integration-variable. Table\,1 shows for the six possible concatenations of two nucleon-lines the proper assignments of the momenta $\vec l_1,\vec l_2,\vec l_3$,  where  $\vec l$ denotes now the unconstrained loop-momentum  and $\vec l_4$ is from the inside of a Fermi sphere, $|\vec l_4|<k_f$. In this setup the angular part of the 3-dimensional loop-integration can be performed analytically, such that all ring contributions to $V_\text{med}$ get reduced to radial integrals $\int_0^\infty\!dl$.
 
 \begin{table}[h!]
\begin{center}
\begin{tabular}{|c|c|c|c|c|c|c|}
    \hline
concat. & $N_3$ on $N_2$  & $N_2$ on $N_3$ &  $N_3$ on $N_1$  & $N_1$ on $N_3$ &  $N_1$ on $N_2$  & $N_2$ on $N_1$  \\  \hline 

$\vec l_1=$  & $\vec l_4+\vec l$  & $-\vec l_4-\vec l$  & $\vec l-\vec p\,'$ &  $\vec p-\vec l$  & $\vec l-\vec p$ &  $\vec p\,'-\vec l$ \\ 

$\vec l_2=$ & $\vec l-\vec p\,'$  &  $\vec p-\vec l$  & $\vec l_4+\vec l$ & $-\vec l_4-\vec l$ & $\vec l-\vec p\,'$ &  $\vec p-\vec l$\\ 

$\vec l_3=$  & $\vec l-\vec p$ &  $\vec p\,'-\vec l$ & $\vec l-\vec p$ & $\vec p\,'-\vec l$ & $\vec l_4+\vec l$ & $-\vec l_4-\vec l$  \\ 
  \hline
  \end{tabular}
\end{center}
{\it Tab.1: Assignment of pion momenta, where  $\vec l$ is unconstrained and 
$|\vec l_4|<k_f$ from a Fermi sphere.}
\end{table}

\subsection{Self-closings of nucleon-lines} 
For the self-closings of nucleon-lines the Fermi sphere integral gives just a factor density $\rho= 2k_f^3/3\pi^2$. Sorted according to powers of $g_A^2$, the self-closing contributions to $V_\text{med}$ from the 3N-ring interaction $V_\text{3N}$ in eq.(35) read: 
\begin{equation}
V^{(0)}_\text{med} = -{g^4_A k_f^3 \over 96\pi^3f_\pi^6}\,\vec\tau_1\!\cdot\!\vec\tau_2\bigg\{{ m_\pi(9m_\pi^2+2q^2)\over 4m_\pi^2+q^2}+{3m_\pi^2+q^2\over q}\arctan{q\over 2m_\pi}\bigg\}\,, \end{equation}
\begin{eqnarray}
V^{(0)}_\text{med}& = &{g^6_A k_f^3\over 96\pi^3f_\pi^6}\bigg\{\vec\tau_1\!\cdot\!\vec\tau_2\bigg[{23 m_\pi^3(2m_\pi^2+q^2)+3m_\pi q^4 \over (4m_\pi^2+q^2)^2}+{3 
m_\pi^2+ q^2\over q}\arctan{q\over 2m_\pi}\bigg] \nonumber\\ &&+ {3\over 2q^2}(\vec\sigma_1\!\cdot\!\vec q\,\vec\sigma_2\!\cdot\!\vec q-\vec\sigma_1\!\cdot\!\vec \sigma_2 q^2 )\bigg[{m_\pi^3\over 4m^2_\pi+q^2}+{q^2-m_\pi^2\over 2q}\arctan{q \over 2m_\pi}\bigg]\bigg\}\,,  \end{eqnarray} 
where we have applied for the isovector central terms $\sim \vec\tau_1\!\cdot\!\vec\tau_2$ the rule of dimensional regularization, $(\int_0^\infty\!dl 1)_\text{dimreg}=0$, to solve the loop integral. This regularization is also inherent in the results for the eleven $R$-functions and seven $S$-functions presented in eqs.(A2,A7) of ref.\,\cite{3Nlong}. 
\subsection{Concatenations of nucleon-lines for ring interaction $\sim g_A^4$}
Treating the concatenations of two nucleon-lines is somewhat simpler for the $g_A^4$-part of the 3N-ring interaction $V_\text{3N}$ in eq.(35), since for this component the three pion-propagators are on an equal footing. Using the assignments of momenta $\vec l_1,\vec l_2,\vec l_3$ for the six possible concatenations in table\,1, one obtains the following contributions to $V_\text{med}$, which we list individually by specifying first their type. 

\noindent Isoscalar central term (one piece):
\begin{eqnarray} V^\text{(cc)}_\text{med} &=& {3g_A^4 \over (4\pi)^4 f_\pi^6} \int_0^\infty\!\!\!dl\bigg\{ l \widetilde\Gamma_1(l)\bigg[\big[ m_\pi^2(8p^2-q^2)+(4p^2+q^2)(p^2-l^2)\big]{\Lambda(l) \over p^2} \nonumber \\ &&+ {l\over p^2}(q^2-4p^2)+2(2m_\pi^2 +q^2)(l^2-p^2-m_\pi^2) \Omega(l)\bigg]+{16k_f^3 \over 3}\bigg\}\,, \end{eqnarray}
with the recurrent auxiliary functions:
\begin{equation} \Lambda(l)={1\over 4p}\ln {m_\pi^2+(l+p)^2 \over m_\pi^2+(l-p)^2}\,,  \end{equation}
\begin{equation} \Omega(l)={1\over q \sqrt{B+q^2l^2}}\ln {q\, l +\sqrt{B+q^2l^2} \over \sqrt{B}}\,,  \end{equation}
and the abbreviation $B= [m_\pi^2+(l+p)^2][m_\pi^2+(l-p)^2]$. In accordance with dimensional regularization the subtraction of the asymptotic constant $-16k_f^3/3$ ensures the convergence of the radial integral $\int_0^\infty \!dl$ in eq.(41).

\noindent Isoscalar and isovector spin-spin and tensor terms (each one piece):
\begin{equation} V^\text{(cc)}_\text{med}={g_A^4(3+\vec\tau_1\!\cdot\!\vec \tau_2)\over 64\pi^4 f_\pi^6}(\vec\sigma_1\!\cdot\!\vec \sigma_2 q^2 -\vec\sigma_1\!\cdot\!\vec q\,\vec\sigma_2\!\cdot\!\vec q\,) \!\int_0^\infty\!\!\!dl{l \widetilde\Gamma_1(l)\over 4p^2-q^2} \bigg[(B+q^2l^2) \Omega(l)-(m_\pi^2+l^2+p^2)\Lambda(l)\bigg]\,, \end{equation}
\noindent Isoscalar and isovector quadratic spin-orbit terms (each one piece):
\begin{eqnarray}
V^\text{(cc)}_\text{med}&=&{g_A^4(3+\vec\tau_1\!\cdot\!\vec \tau_2)\over 64\pi^4 f_\pi^6} \,\vec\sigma_1\!\cdot\!(\vec q \!\times \! \vec p\,)\vec\sigma_2\!\cdot\!(\vec q \!\times \! \vec p\,) \!\int_0^\infty\!\!\!dl{l \widetilde\Gamma_1(l)\over 4p^2-q^2} \bigg\{-{l \over p^2}+ \bigg[{m_\pi^2+l^2+p^2\over p^2}\nonumber \\ && +{2(4m_\pi^2+4l^2+q^2)\over 4p^2 -q^2}\bigg]\Lambda(l) +2\bigg[m_\pi^2+3l^2+p^2 -{4( m_\pi^2+l^2+p^2)^2 \over 4p^2-q^2}\bigg] \Omega(l)\bigg\} \,, \end{eqnarray}
\noindent Isovector central term (two pieces added):
\begin{eqnarray}
V^\text{(cc)}_\text{med}&=&{g_A^4\vec\tau_1\!\cdot\!\vec \tau_2\over 64\pi^4 f_\pi^6} \!\int_0^\infty\!\!\!dl \bigg\{ \big[3l\Gamma_2(l)+l^3 \widetilde\Gamma_3(l)\big] \big[(2m_\pi^2+q^2)\Omega(l)-2\Lambda(l) \big] +l \widetilde\Gamma_1(l)\bigg[ {l \over 4p^2}(q^2-4p^2) \nonumber \\ &&+\Big[2m_\pi^2+p^2-l^2 +{q^2 \over 4} -{q^2 \over 4p^2}(m_\pi^2\!+\!l^2)\Big]\Lambda(l)+\Big( m_\pi^2\!+\!{q^2\over 2}\Big)(l^2\!-\!p^2\!-\!m_\pi^2) \Omega(l)\bigg]+{8k_f^3 \over 3} \bigg\} \,, \nonumber \\ \end{eqnarray}
\noindent Isovector spin-orbit term (three pieces added):
\begin{eqnarray}
V^\text{(cc)}_\text{med}&=&{g_A^4\vec\tau_1\!\cdot\!\vec \tau_2\over 64\pi^4 f_\pi^6}\, i(\vec\sigma_1\!+\!\vec\sigma_2)\!\cdot\!(\vec q \!\times \! \vec p\,)\!\int_0^\infty\!\!\!dl{l \over 4p^2-q^2}\bigg\{ \widetilde\Gamma_1(l)\Big[ (p^2+l^2) \Lambda(l)-\big[m_\pi^2(p^2+l^2)\nonumber \\ &&+q^2l^2+(p^2-l^2)^2\big] \Omega(l)\Big]+ 4\Gamma_2(l)\bigg[\Big(m_\pi^2+l^2-p^2+{q^2\over 2}\Big)\Omega(l)- 
\Lambda(l)\bigg]\nonumber \\ && +{\widetilde\Gamma_3(l)\over 2} \Big[ 
\big[(m_\pi^2\!+\!p^2)^2+3l^4+2l^2(2m_\pi^2\!-\!2p^2\!+\!q^2)\big]  \Omega(l)-(m_\pi^2\!+\!3l^2\!+\!p^2) \Lambda(l)\Big] \bigg\}\,.  \end{eqnarray}
The analytical expressions for the functions $\widetilde\Gamma_1(l)$,  $\Gamma_2(l)$ and $\widetilde\Gamma_3(l)$ are given in the appendix.  

\subsection{Concatenations of nucleon-lines for ring interaction $\sim g_A^6$}
When treating the concatenations of two nucleon-lines for the $g_A^6$-part of the 3N-ring interaction $V_\text{3N}$, one has to distinguish the cases with 
$\vec l_2 =\pm(\vec l_4+\vec l\,)$, because this momentum enters a squared pion-propagator in eq.(35). The Fermi sphere integral over the squared pion-propagator combined with the solution of the angular integral leads to the following contributions to $V_\text{med}$ from the concatenations $N_1$ on $N_3$ and $N_3$ on $N_1$.

\noindent Isoscalar central term:
\begin{eqnarray} V^\text{(2)}_\text{med} &=& {3g_A^6 \over (4\pi)^4 f_\pi^6} \int_0^\infty\!\!\!dl\bigg\{ 4l \gamma_2(l)\bigg[{l\over p^2}(4p^2\!-\!q^2)+\Big((m_\pi^2+l^2) {q^2\over p^2} -8m_\pi^2-3q^2\Big)\Lambda(l)\nonumber \\ &&+ (2m_\pi^2 + q^2)^2\, \Omega(l)\bigg]+ {l\over 2} \widetilde \gamma_3(l)\bigg[
{l\over p^2} \big[ 2p^2(7l^2-2m_\pi^2-p^2)+q^2(m_\pi^2-3l^2)\big]  +\Big[ 2p^4\nonumber \\ &&+p^2(8m_\pi^2-4l^2+q^2)+6m_\pi^2(m_\pi^2-4l^2)+2l^2(l^2-4q^2) +{q^2 \over p^2}(3l^4+2m_\pi^2l^2-m_\pi^4) \Big] \Lambda(l)\nonumber \\ && +(2m_\pi^2+q^2) \big[2l^2(3m_\pi^2+p^2+q^2)-l^4-(m_\pi^2+p^2)^2\big] \Omega(l)\bigg] -{16k_f^3 \over 3}\bigg\}\,, \end{eqnarray}
\noindent Isoscalar spin-orbit term:
\begin{eqnarray} V^\text{(2)}_\text{med} &=& {3g_A^6 \over (4\pi)^4 f_\pi^6} i (\vec\sigma_1\!+\!\vec\sigma_2)\!\cdot\!(\vec q \!\times \! \vec p\,)\!\int_0^\infty\!\!\!dl{l \over 4p^2-q^2}\bigg\{\gamma_2(l)\bigg[\Big[ 4(p^2\!-\!l^2\!-\!2m_\pi^2)-3q^2 +{q^2\over p^2}(m_\pi^2\!+\!l^2)\Big]\Lambda(l) \nonumber \\ && +{l \over p^2}(4p^2\!-\!q^2) +(2m_\pi^2\!+\!q^2)(2m_\pi^2+2l^2-2p^2+q^2)\Omega(l)\bigg] +\widetilde\gamma_3(l)\bigg[ {l \over 4p^2}(4p^2\!-\!q^2)(m_\pi^2\!+\!l^2\!+p^2) \nonumber \\ && +\Big[ p^2\Big( 2l^2\!-4m_\pi^2-{3q^2\over 4} -p^2\Big)+{q^2 \over 4p^2}(m_\pi^2\!+\!l^2)^2 -l^4-l^2\Big(4m_\pi^2\!+\!{3q^2\over2} \Big) 
-m_\pi^2\Big(3m_\pi^2\!+\!{q^2\over2} \Big)\Big] \Lambda(l) \nonumber \\ && +(
2m_\pi^2+q^2)\big[l^4+l^2(2m_\pi^2-2p^2+q^2)+(m_\pi^2+p^2)^2\big] \Omega(l)\bigg] 
\bigg\}\,, \end{eqnarray}
\noindent Isovector spin-orbit term:
\begin{eqnarray} V^\text{(2)}_\text{med} &=& {g_A^6 \vec\tau_1\!\cdot\!\vec \tau_2\over  128\pi^4 f_\pi^6} i (\vec\sigma_1\!+\!\vec\sigma_2)\!\cdot\!(\vec q \!\times \! \vec p\,)\!\int_0^\infty\!\!\!dl{l \over 4p^2-q^2}\bigg\{ \gamma_2(l)\bigg[\Big[ 4(2m_\pi^2\!+\!l^2\!-\!p^2) +3q^2 -{q^2\over p^2}(m_\pi^2\!+\!l^2)\Big]\Lambda(l)  \nonumber \\ && +{l \over p^2}(q^2-4p^2)+(2m_\pi^2+q^2)(2p^2-2m_\pi^2-2l^2-q^2)\Omega(l) \bigg] +\widetilde\gamma_3(l)\bigg[ {l \over 2p^2}(4p^2-q^2)(m_\pi^2-l^2) \nonumber \\ && +(m_\pi^2+p^2-l^2) \Big[p^2-l^2-3m_\pi^2 -{q^2 \over 2}+{q^2 \over 2p^2}(m_\pi^2+l^2)\Big] \Lambda(l)+(m_\pi^2+p^2-l^2)^2 \nonumber \\ && \times \Big(m_\pi^2+l^2-p^2+{q^2\over 2}\Big) \Omega(l)\bigg] \bigg\}\,, \end{eqnarray}
\noindent Isovector spin-spin and tensor terms:
\begin{eqnarray} V^\text{(2)}_\text{med} &=& {g_A^6 \vec\tau_1\!\cdot\!\vec \tau_2\over  64\pi^4 f_\pi^6}(\vec\sigma_1\!\cdot\!\vec \sigma_2 q^2 -\vec\sigma_1\!\cdot\!\vec q\,\vec\sigma_2\!\cdot\!\vec q\,) \!\int_0^\infty\!\!\!dl{l \over 4p^2-q^2}\bigg\{ 4\gamma_2(l) \Big[(m_\pi^2+l^2+p^2) \Lambda(l)-(B+q^2l^2)  \Omega(l)\Big] \nonumber \\ && + \widetilde\gamma_3(l) \bigg[{l \over 8p^2} (4p^2-q^2)(m_\pi^2+l^2+p^2)+\Big[ p^2\Big(l^2-3m_\pi^2+{q^2 \over 8}-{3p^2 \over2} \Big)+{l^4 \over 2} -m_\pi^2l^2 -{3m_\pi^4\over2} \nonumber \\ &&+{q^2 \over 4}(m_\pi^2-l^2)+{q^2 \over 8p^2}(m_\pi^2+l^2)^2\Big] \Lambda(l) +(m_\pi^2-l^2+p^2) (B+q^2l^2)  \Omega(l)\bigg]\bigg\}\,, \end{eqnarray}
\noindent Isovector quadratic spin-orbit term:
\begin{eqnarray}
V^\text{(2)}_\text{med}&=&{g_A^6 \vec\tau_1\!\cdot\!\vec \tau_2\over 64\pi^4 f_\pi^6} \,\vec\sigma_1\!\cdot\!(\vec q \!\times \! \vec p\,)\vec\sigma_2\!\cdot\!(\vec q \!\times \! \vec p\,) \!\int_0^\infty\!\!\!dl{l\over 4p^2-q^2} \bigg\{4\gamma_2(l)\bigg[ {l \over p^2}+\Big[1-{ m_\pi^2+l^2\over p^2} \nonumber \\ &&-{2(4m_\pi^2+4l^2+q^2)\over  4p^2-q^2} \Big] \Lambda(l) +\Big[ {8(m_\pi^2+l^2+p^2)^2 \over 4p^2-q^2} -8l^2-4m_\pi^2-q^2\Big] \Omega(l) \bigg]\nonumber \\ &&  + \widetilde\gamma_3(l)\bigg[{l \over p^2}(l^2-m_\pi^2-p^2) +\Big[ 
{m_\pi^4-l^4\over p^2} +2l^2 -p^2 +{8 \over 4p^2-q^2}\big((m_\pi^2+p^2)^2-l^4\big) \Big] \Lambda(l) \nonumber \\ && +2(l^2-m_\pi^2 -p^2)\Big[{4(m_\pi^2+l^2+ p^2)^2 \over 4p^2-q^2}-3l^2-m_\pi^2-p^2\Big] \Omega(l)\bigg]\bigg\}\,, \end{eqnarray}
with the explicit expressions for the functions  $\gamma_2(l)$ and $\widetilde \gamma_3(l)$ given in the appendix. 

For the other four concatenations one has $\vec l_{1,3} = \pm(\vec l_4+\vec l\,)$ and the Fermi sphere integral goes over an ordinary pion-propagator. Actually, the lengthy formulas simplify when adding two pieces of the same type, such that one obtains the following additional contributions to $V_\text{med}$ from concatenations. 

\noindent Isoscalar spin-spin and tensor term:
\begin{eqnarray} V^\text{(cc)}_\text{med} &=& {3g_A^6 \over (4\pi)^4 f_\pi^6}(\vec\sigma_1\!\cdot\!\vec \sigma_2 q^2 -\vec\sigma_1\!\cdot\!\vec q\,\vec\sigma_2\!\cdot\!\vec q\,) \!\int_0^\infty\!\!\!dl{l \over 4p^2-q^2}\bigg\{ 12\Gamma_2(l)\Big[(m_\pi^2+l^2+ p^2)\Omega(l)-\Lambda(l)\Big] \nonumber \\ && +\widetilde\Gamma_3(l)\bigg[{l \over 2p^2}(q^2-4p^2)+\Big[3(3m_\pi^2+l^2+ 3p^2)-{q^2 \over 2} -{q^2 \over 2p^2}(m_\pi^2+l^2)\Big] \Lambda(l) \nonumber \\ && - \Big[ 4l^2(2m_\pi^2+q^2-2p^2)+l^4 +7(m_\pi^2+p^2)^2\Big] \Omega(l)\bigg] \bigg\}\,, \end{eqnarray}
\noindent Isoscalar quadratic spin-orbit term:
\begin{eqnarray}
V^\text{(cc)}_\text{med}&=&{3g_A^6 \over 64\pi^4 f_\pi^6} \,\vec\sigma_1\!\cdot\!(\vec q \!\times \! \vec p\,)\vec\sigma_2\!\cdot\!(\vec q \!\times \! \vec p\,) \!\int_0^\infty\!\!\!dl{1\over 4p^2-q^2} \bigg\{3\Gamma_2(l)\bigg\{\bigg( {l\over p^2}+{8l \over 4p^2-q^2}\bigg)\Lambda(l)-{1\over p^2} \nonumber \\ && +{m_\pi^2 +l^2+p^2\over q^2}\Big[  {m_\pi^2(q^2-4p^2)\over p^2 B}+{4m_\pi^2+q^2\over B+q^2l^2} \Big]- \bigg[{4m_\pi^2+q^2\over B+q^2l^2}(m_\pi^2 +l^2+p^2)+{4 \over 4p^2-q^2}\nonumber \\ && \times (2m_\pi^2 +2l^2-2p^2+q^2)\bigg] l\, \Omega(l) \bigg\} + l \widetilde\Gamma_3(l)\bigg\{-2\bigg({ m_\pi^2\over p^2} +{7m_\pi^2+ 7p^2+l^2 \over 4p^2-q^2}\bigg) \Lambda(l) \nonumber \\ && +{l\over q^2}+{l \over 2q^2}(m_\pi^2+l^2 + p^2)\Big[{4m_\pi^2 (q^2-4p^2)\over p^2 B} +{3(4m_\pi^2+q^2) \over B+q^2l^2}\Big] \nonumber \\ &&+\bigg[{m_\pi^4-(l^2-p^2)^2 \over q^2} +{2\over 4p^2-q^2} \Big(l^4+8l^2( m_\pi^2+ p^2)+7(m_\pi^2+ p^2)^2\Big) \nonumber \\ &&-{1\over 2}(3m_\pi^2+7l^2+3p^2) -{3l^2 \over B+q^2l^2} \Big(2m_\pi^2+{q^2\over 2}\Big)(m_\pi^2+l^2+p^2)  \bigg] \Omega(l)\bigg\} \bigg\}\,,\end{eqnarray}
\noindent Isovector central term: 
\begin{eqnarray}
V^\text{(cc)}_\text{med}&=&{g_A^6\vec\tau_1\!\cdot\!\vec \tau_2\over 128\pi^4 f_\pi^6} \!\int_0^\infty\!\!\!dl \bigg\{ \big[3\Gamma_2(l)+l^2 \widetilde\Gamma_3(l)\big] \bigg\{  {q^2 \over p^2}-{(2m_\pi^2 +q^2)^2 \over q^2(B+q^2l^2)} (m_\pi^2+ l^2+p^2)\nonumber \\ && +{ m_\pi^2  \over B}\bigg[4(m_\pi^2+p^2-l^2) +{4m_\pi^2 p^2-q^4 \over p^2q^2}(m_\pi^2+l^2+p^2)\bigg] +{l\over p^2}(8p^2-q^2) \Lambda(l) \nonumber \\ &&+\bigg[{2m_\pi^2 +q^2\over B+q^2l^2} (m_\pi^2+l^2+p^2) -4\bigg]
(2m_\pi^2 +q^2)l\,\Omega(l) \bigg\}-{16k_f^3\over 3}\bigg\}\,,  \end{eqnarray}
\noindent Isovector spin-orbit  term:
\begin{eqnarray} V^\text{(cc)}_\text{med} &=& {g_A^6 \vec\tau_1\!\cdot\!\vec \tau_2\over  (4\pi)^4 f_\pi^6} i (\vec\sigma_1\!+\!\vec\sigma_2)\!\cdot\!(\vec q \!\times \! \vec p\,)\!\int_0^\infty\!\!\!dl\,\bigg\{4\Gamma_2(l)\bigg\{\bigg[{l\over p^2}+{4l \over 4p^2-q^2}\bigg] \Lambda(l)-{1\over p^2}\nonumber \\ &&+ {m_\pi^2 \over B}\bigg[ 1+{m_\pi^2+l^2\over p^2}+{2\over q^2}(l^2-m_\pi^2-p^2)\bigg]+{2m_\pi^2+q^2\over q^2(B+q^2l^2)} (m_\pi^2+p^2-l^2) \nonumber \\ &&+\bigg[2-{4(m_\pi^2 +l^2+p^2)\over 4p^2-q^2} +{2m_\pi^2+q^2\over B+q^2l^2}(l^2-m_\pi^2-p^2)\bigg] l\, \Omega(l) \bigg\} \nonumber \\ && +l \widetilde\Gamma_3(l)\bigg\{ 
l(m_\pi^2+p^2-l^2) \bigg[ {2m_\pi^2+q^2\over q^2(B+q^2l^2)}+{1\over B}\Big( {m_\pi^2+l^2 \over p^2}-{4m_\pi^2\over q^2}-1\Big)\bigg] \nonumber \\ && +\bigg[{l^2-m_\pi^2- p^2 \over p^2} +{2(3l^2+m_\pi^2+p^2) \over 4p^2-q^2}\bigg] \Lambda(l)
+\bigg[(m_\pi^2+p^2-l^2)\nonumber \\ && \times \bigg({2m_\pi^2\over q^2}-{(2m_\pi^2+q^2) l^2\over B+q^2l^2} \bigg) +{ m_\pi^2+p^2+3l^2\over 4p^2-q^2}\big(2p^2-2m_\pi^2-2l^2-q^2\big) \bigg]\Omega(l) \bigg\} \bigg\}\,. \end{eqnarray}
We note as an aside that for the isovector central terms $\sim\vec\tau_1\!\cdot\!\vec \tau_2$ in eqs.(39,40,46,55) the subtraction constants $\sim k_f^3$ cancel each other. Therefore, the linear divergences appearing in cutoff regularization for the isoscalar central terms in eqs.(41,48) could be absorbed into the short-distance parameter $c_E$, which produces just a spin- and isospin-independent $V_\text{med}$ linear in density $\rho$ (see eq.(25) in ref.\,\cite{holt}).
   
Before closing this chapter, we consider as an outlook to future work the simplest
3N-ring diagram at N$^4$LO. It is proportional to $c_{1,2,3,4}$ and independent of $g_A$, with the corresponding coordinate-space potential written in eq.(4.9) of ref.\,\cite{midrange4}. Working with the 3N-interaction in momentum-space, the self-closing of nucleon-lines produces an isovector central in-medium potential of the form:
\begin{eqnarray} V^{(0)}_\text{med} &=& {k_f^3 \,\vec\tau_1\!\cdot\!\vec \tau_2\over  48\pi^4 f_\pi^6} \bigg\{ \xi\bigg[\Big( 2c_1\!-\!{3c_2\over 2} \!-\!3c_3\Big)m_\pi^2 -\Big({c_2\over 4} +{c_3\over 3}\Big) q^2\bigg] +\Big( {7c_2\over 4}\!-\!2c_1\!+\!{10c_3\over 3} \Big)m_\pi^2+(3c_2+4c_3){q^2 \over 9} \nonumber \\ && +\bigg[\Big( 2c_1-c_2 -{7c_3\over 3}\Big){m_\pi^2\over q}- \Big({c_2\over 4} +{c_3\over 3}\Big) q  \bigg]\sqrt{4m_\pi^2+q^2}\ln{q+\sqrt{4m_\pi^2+q^2}\over 
2m_\pi}\bigg\}\,,  \end{eqnarray} 
where $\xi=1/(d-4)+(\gamma_E -\ln4\pi)/2 +\ln(m_\pi/\lambda)$ denotes the ultraviolet divergence in dimensional regularization with some scale $\lambda$. Together with further ultraviolet divergences, it can be absorbed on the short-distance parameters $c_E$ and $E_{1,...,10}$ \cite{3Ncontact} (see also eq.(49) in ref.\,\cite{vmedshort}). Note that the combination $c_2/4 +c_3/3$ appears twice such that the chiral limit $m_\pi\to 0$ of $V^{(0)}_\text{med}$ exists. Alternative regularizations (by a 3-dimensional or 4-dimensional cutoff) leave the terms with logarithms in eq.(57) intact, but change the purely polynomial part. 
\section{Two-pion exchange three-nucleon force at N$^4$LO}
In this section we treat the longest-range $2\pi$-exchange 3N-interaction following the work of ref.\cite{twopi4}. Modulo terms of shorter range it can be written according to eq.(3.1) in ref.\cite{twopi4} in the general form:
\begin{equation}2V_\text{3N}={g_A^2 \over 4f_\pi^4} {\vec\sigma_1\!\cdot\! \vec
q_1 \vec \sigma_3\!\cdot\! \vec q_3 \over (m_\pi^2+q_1^2) (m_\pi^2+q_3^2)}\big[
\vec\tau_1 \!\cdot\! \vec\tau_3\,\tilde g_+(q_2) +\vec\tau_1\!\cdot\!(\vec\tau_3
\!\times\! \vec\tau_2)\, \vec\sigma_2\!\cdot\!(\vec q_1\!\times\!\vec q_3)\,
\tilde h_-(q_2)\big]\,, \end{equation}
with both sides multiplied again by a factor $2$ due to the $1\!\leftrightarrow\!3$ symmetry. In our modified notation (with a prefactor $g_A^2/4f_\pi^4$ taken out) the two structure functions $\tilde g_+(q_2)$ and $\tilde h_-(q_2)$ are $f_\pi^2$ times the isoscalar non-spinflip and isovector spinflip $\pi N$-scattering amplitude at zero pion-energy $\omega=0$ and squared momentum-transfer $t= - q_2^2$. This input\footnote{Since they originate from on-shell $\pi N$-scattering amplitudes, the structure functions $\tilde g_+(q_2)$ and $\tilde h_-(q_2)$ could actually be determined in $\pi N$-dispersion relation analyses from data, without invoking chiral effective field theory.} to the $2\pi$-exchange 3N-interaction is given (and calculated) order by order in chiral perturbation theory and according to eqs.(3.3,3.4,3.14) in ref.\cite{twopi4} these structure functions read up to N$^4$LO:
\begin{eqnarray}\tilde g_+(s)&=&(2c_3-4c_1)m_\pi^2+c_3 s^2 \nonumber \\ &&
+{g_A^2 \over 32 \pi
f_\pi^2} \big[ (2m_\pi^4+5m_\pi^2 s^2+2s^4) A(s) +(4g_A^2+1)m_\pi^3+2(g_A^2+1)
m_\pi s^2 \big]\nonumber \\ &&+ {1\over (4\pi f_\pi)^2} \big[ (c_1'm_\pi^2+c_2'
s^2)(m_\pi^2+2s^2) L(s) +d_1 m_\pi^4+ d_2 m_\pi^2 s^2+ d_3 s^4\big] \,,
\end{eqnarray}
\begin{eqnarray}\tilde h_-(s)&=& c_4 -{g_A^2 \over 32 \pi
f_\pi^2} \big[ (4m_\pi^2+s^2) A(s) +(2g_A^2+1)m_\pi \big] \nonumber \\ &&+ {1\over (4\pi f_\pi)^2} \Big[ {c_4\over 3 }(4m_\pi^2+s^2) L(s) +d_4 m_\pi^2+ d_5s^2 \Big] \,. \end{eqnarray}
Here, we have introduced the combinations  $c_1' = 2c_3-4c_1+2c_2/3$ and $c_2' = c_3+c_2/6$ of low-energy constants and the logarithmic loop function:  
\begin{equation}L(s) ={\sqrt{4m_\pi^2+s^2} \over s} \ln {s+ \sqrt{4m_\pi^2 +s^2}\over 2m_\pi} \,, \end{equation}
while $d_1,\dots, d_5$ abbreviate lengthy combinations of parameters in eq.(3.14) of ref.\cite{twopi4}. It is worth to note that the general form in eq.(58) covers also the recent work \cite{twopidelta} on the $2\pi$-exchange 3N-interaction in chiral effective field theory with explicit 
$\Delta(1232)$-isobars. The pertinent structure functions  $\tilde g_+(q_2)$ and  $\tilde h_-(q_2)$ are given by $8f_\pi^4/g_A^2$ times the expressions written in eq.(5.4) of ref.\,\cite{twopidelta}. The latter involve the loop function $L(q_2)$, polynomial pieces with specific parameters $d_1, d_2, d_4$, and the extra loop function 
\begin{equation} D(q_2) = {1\over \Delta} \int_{2m_\pi}^\infty\!d\mu \, {1\over \mu^2 +q_2^2} \arctan{\sqrt{\mu^2-4m_\pi^2} \over 2\Delta}\,, \end{equation}
which arises from propagating $\Delta(1232)$-isobars, where $\Delta= 293\,$MeV denotes the $\Delta N$-mass splitting. The leading terms from isobar-excitation are obtained by setting $c_1= 0$, $ c_3= -g_A^2/2\Delta$, $c_4 = g_A^2/4\Delta $.

\subsection{Contributions to in-medium NN-potential}
Here, we calculate the contributions to the in-medium NN-potential $V_\text{med}$ from the general $2\pi$-exchange 3N-interaction in eq.(58). The self-closing of nucleon-line $2$ gives (after relabeling $3\to 2$) the contribution:
\begin{equation}V^{(0)}_\text{med} = {g^2_A m_\pi^2 k^3_f \over 3\pi^2 f^4_\pi}
{\vec\tau_1\!\cdot\!\vec\tau_2\over (m^2_\pi+q^2)^2} \,\vec\sigma_1\!\cdot\!\vec
q\, \vec\sigma_2\!\cdot\!\vec q\, \Big\{ 2c_1-c_3 -{m_\pi \over 16 \pi f_\pi^2} 
\Big[ g_A^4+{3g_A^2 \over 8} +{m_\pi \over 2\pi}(c_1'+d_1)\Big]\Big\}\,, \end{equation}
where the value in parenthesis is $-\tilde g_+(0)/2m_\pi^2$. From pionic vertex corrections on either nucleon-line one obtains the (total) contribution:
\begin{eqnarray} V^{(1)}_\text{med} & = &{g^2_A\over 16\pi^2 f^4_\pi} {\vec\tau_1\!\cdot\!\vec\tau_2\over m^2_\pi+q^2} \,\vec\sigma_1\!\cdot\!\vec
q\, \vec\sigma_2\!\cdot\!\vec q\int_{p-k_f}^{p+k_f}\!\! ds\, {s\over q^3} \bigg\{ \Big[(q^2-m_\pi^2-s^2)\tilde g_+(s)\nonumber \\ && -\big(m_\pi^2+(q+s)^2\big) \big(m_\pi^2+(q-s)^2\big)\tilde h_-(s)\Big]  \ln{ q X + 2\sqrt{W} \over (2p+q)[m_\pi^2+(q-s)^2]} \nonumber \\ &&+ {q\over p} \big[k_f^2-(p-s)^2\big] \Big[\tilde g_+(s)+ \Big(m_\pi^2 +s^2+ {q^2\over 4}\Big(5 +{2s \over p}+{s^2-k_f^2 \over p^2}\Big)\Big)\tilde h_-(s)\Big]\bigg\}\,,  \end{eqnarray}
with the auxiliary polynomials $X$ and $W$ defined in eq.(9). At second order one has in $\tilde g_+(s)$ and $\tilde h_-(s)$ only the low-energy constants $c_{1,3,4}$ and the corresponding integral $\int_{p-k_f}^{p+k_f}\! ds...$ in eq.(64) is solved by:
\begin{equation}2(c_3-4c_1)m_\pi^2\widetilde\Gamma_1 +2(c_3+c_4) \bigg[ {2k_f^3 \over 3} -q^2(\Gamma_1+\Gamma_3)-4\Gamma_2\bigg] +2c_4\big[ 2k_f^3-(4m_\pi^2+q^2)\Gamma_0-q^2 \Gamma_1\big]\,, \end{equation}
which serves as a good (numerical) check. In the same way, the purely polynomial amplitudes proportional to $d_{1,2,3,4,5}$ at fourth order lead for the integral 
$\int_{p-k_f}^{p+k_f}\! ds...$ in eq.(64) to the result: 
\begin{eqnarray}&& {1\over (2\pi f_\pi)^2}\bigg\{m_\pi^2\big[(d_1-d_2+d_3)m_\pi^2+(d_5-d_4)q^2\big]{\widetilde\Gamma_1 \over 2}+\big[(d_2-2d_3+d_4-d_5)m_\pi^2-d_5 q^2\big]\nonumber \\ && \times \bigg[{k_f^3\over 3}-{q^2\over 2}(\Gamma_1+\Gamma_3) -2\Gamma_2 \bigg] +(d_3+d_5)\bigg[k_f^3\Big({m_\pi^2+p^2\over 3}-{k_f^2\over 5}\Big)+2q^2(\Gamma_2+3\Gamma_4)+{q^4\over 2}(\Gamma_3+\Gamma_5)\bigg]\nonumber \\ && +d_4 m_\pi^2(k_f^3-2m_\pi^2\Gamma_0)+d_5\bigg[ k_f^3\Big(p^2-{4m_\pi^2\over 3}+{11k_f^2 \over 15}\Big) +2m_\pi^2(m_\pi^2\Gamma_0+q^2 \Gamma_1)\bigg]\bigg\}\,,  \end{eqnarray}
which also serves as a good (numerical) check. Finally, the two diagrams related to double exchange lead to the expression: 
\begin{eqnarray} V^{(3)}_\text{med} & = &{g^2_A\over 16\pi^2 f^4_\pi} \Big\{
3 \tilde g_+(q)\big[ 2\Gamma_0-(2m_\pi^2+q^2) G_0\big] + i( \vec\sigma_1\!+\!\vec\sigma_2)\!\cdot\!(\vec q\!\times \! \vec p\,)\Big[3 \tilde g_+(q)(G_0+2G_1) \nonumber \\ &&+\vec\tau_1\!\cdot\!\vec\tau_2\, \tilde h_-(q)\big[ (2m_\pi^2+q^2) (G_0+2G_1) -2\Gamma_0-2\Gamma_1\big] \Big]+4\vec\tau_1\!\cdot\!\vec\tau_2\, \tilde h_-(q)\nonumber \\ && \times \Big[ (\vec\sigma_1\!\cdot\!\vec \sigma_2 q^2 -\vec\sigma_1\!\cdot\!\vec q\, \vec\sigma_2\!\cdot\!\vec q\,)G_2+ \vec\sigma_1\!\cdot\!(\vec q\! \times\! \vec p\,)\vec \sigma_2\!\cdot\!(\vec q\!\times\!\vec p\,)(G_0+
4G_1 +4G_3)\Big] \Big\}\,,  \end{eqnarray}
where we remind that the functions $\Gamma_\nu$ depend on $(p,k_f)$, and the functions $G_\nu$ depend on $(p,q,k_f)$. 

\subsection{Relativistic $1/M$-correction}
In the formulation of chiral effective field theory with explicit $\Delta(1232)$-isobar degrees of freedom also the first relativistic $1/M$-correction to the $2\pi$-exchange 3N-interaction has been derived. Choosing the large-$N_c$ value for the $\pi N\Delta$-coupling constant $h_A = 3\sqrt{2}g_A/4$, this 3N-interaction 
term of order N$^3$LO reads according to eq.(5.16) in ref.\,\cite{twopidelta} as:
\begin{eqnarray} 2V_\text{3N} &=&  {g_A^4 \over 64 Mf_\pi^4 \Delta^2} {\vec\sigma_1\!\cdot\! \vec q_1 \vec \sigma_3\!\cdot\! \vec q_3 \over (m_\pi^2+q_1^2) (m_\pi^2+q_3^2)}\Big\{\vec\tau_1 \!\cdot\! \vec\tau_3\Big[-8(\vec q_1\!\cdot\!\vec q_3)^2 \nonumber \\ && + i \vec\sigma_2\!\cdot\!(\vec q_1\!\times\!\vec q_3)\big((\vec p_1+\vec p_1\,\!\!')\!\cdot\!(2 \vec q_1-\vec q_3) +(\vec p_3+\vec p_3\,\!\!')\!\cdot\!( \vec q_1-2\vec q_3)\big)\Big]  \nonumber \\ && + i\vec\tau_1\!\cdot\!(\vec\tau_2 \!\times\! \vec\tau_3)\vec q_1\!\cdot\!\vec q_3\Big[ (\vec p_1+\vec p_1\,\!\!')\!\cdot\!(\vec q_3-2\vec q_1)+(\vec p_3+\vec p_3\,\!\!')\!\cdot\!(2 \vec q_3-\vec q_1)+2i \vec\sigma_2\!\cdot\!(\vec q_1\!\times\!\vec q_3)\Big]\Big\}\,,\end{eqnarray}
and it has the interesting property of scaling with the inverse square of the small scale $\Delta=293\,$MeV. Due to the four powers of momenta in the curly bracket, this $V_\text{3N}$ is not expressible through the $1/M$-corrections to the $2\pi$-exchange 3N-force treated in subsection V.B of ref.\cite{vmedshort}. Therefore, we have to calculate once more its contributions to $V_\text{med}$.  
The self-closing of nucleon-line 2 gives (after relabeling $3\to 2$) the piece:
\begin{equation}V_\text{med}^{(0)} = {g_A^4 k_f^3\, q^4\over 12\pi^2 Mf_\pi^4 \Delta^2} {\vec\tau_1 \!\cdot\! \vec\tau_2 \over  (m_\pi^2+q^2)^2} \, \vec \sigma_1\!\cdot\! \vec q\, \vec \sigma_2\!\cdot\! \vec q\,,  \end{equation} 
which comes entirely from the first term porportional to $(\vec q_1\!\cdot\!\vec q_3)^2$ in eq.(68).  From pionic vertex corrections on either nucleon-line one obtains the following (total) contribution:
\begin{eqnarray} V_\text{med}^{(1)} &=&{g_A^4 \vec\tau_1 \!\cdot\! \vec\tau_2 \over 32\pi^2 Mf_\pi^4 \Delta^2} {\vec \sigma_1\!\cdot\! \vec q\, \vec \sigma_2\!\cdot\! \vec q \over m_\pi^2+q^2} \bigg\{ {8k_f^5\over 15} -\Big(4m_\pi^2+p^2 +{31q^2\over 4}\Big)\Gamma_2 -\Big(p^2 +{7q^2 \over 2}\Big) \Gamma_4 -m_\pi^2 p^2(\Gamma_1+ \Gamma_3) \nonumber \\ && +{m_\pi^2 q^2 \over 4}(2 \Gamma_1 -\Gamma_3)  - {q^2 \over 4}(p^2+q^2)\big(6\Gamma_0+13 \Gamma_1 +8\Gamma_3+\Gamma_5\big)  +{9q^4 \over 8}(\Gamma_0+2\Gamma_1 +\Gamma_3) \bigg\}\,, \end{eqnarray}
with the functions $\Gamma_{4,5}$ given in the appendix of ref.\,\cite{vmedshort}. For the contributions to  $V_\text{med}$ from double exchange we present, in view of the lengthy formulas, the isoscalar and isovector parts separately:
\begin{eqnarray} V_\text{med}^{(3)} &=&{3g_A^4 \over 64\pi^2 Mf_\pi^4 \Delta^2} \bigg\{{4k_f^3\over 3}\Big(4m_\pi^2-2p^2+3q^2-{6k_f^2 \over 5}\Big) -2q^2(2\Gamma_2+3m_\pi^2 \Gamma_1) -q^4(3\Gamma_1+\Gamma_3) \nonumber \\ && 
-3(2m_\pi^2+q^2)^2 \Gamma_0 +(2m_\pi^2+q^2)^3 {G_0\over 2}+ (\vec \sigma_1\!\cdot\!\vec \sigma_2 q^2-\vec \sigma_1\!\cdot\! \vec q\, \vec \sigma_2\!\cdot\! \vec q\,)  \Big[ (6p^2-q^2){ G_2\over 2}-3G_{2*}\Big] \nonumber \\ && +\vec\sigma_1\!\cdot\!(\vec q\! \times\! \vec p\,)\vec \sigma_2\!\cdot\!(\vec q\!\times\!\vec p\,)\bigg[(6p^2-q^2)\Big({ G_0\over 2}+2G_1 +2G_3\Big)-3(G_{0*}+4G_{1*} +4G_{3*})\bigg]\nonumber \\ && + i( \vec\sigma_1\!+\!\vec\sigma_2)\!\cdot\!(\vec q\!\times \! \vec p\,)\bigg[4m_\pi^2(\Gamma_0+\Gamma_1)-{4k_f^3 \over 3}+{3p^2\over 2} (\Gamma_3+\Gamma_5-\Gamma_0-\Gamma_1) +{3\over 2}(3\Gamma_2+5\Gamma_4)\nonumber \\ && +{q^2\over 4} (9\Gamma_0+13\Gamma_1+4\Gamma_3)-(2m_\pi^2+q^2)\bigg({3\over 4}(G_{0*}\!+\!2G_{1*}) +\Big(m_\pi^2\!-\!{3p^2\over 4}\!+\!{5q^2 \over 8}\Big)(G_0\!+\!2G_1)\bigg) \bigg] \bigg\}\,, \nonumber \\ \end{eqnarray}

\begin{eqnarray} V_\text{med}^{(3)} &=&{g_A^4 \vec\tau_1 \!\cdot\! \vec\tau_2 \over 64\pi^2 Mf_\pi^4 \Delta^2} \bigg\{4k_f^3\Big(p^2-{q^2\over 6}-{3k_f^2 \over 5}\Big) +{q^2\over 2} (30 \Gamma_4+q^2 \Gamma_1)+3p^2q^2(\Gamma_5-  \Gamma_1) \nonumber \\ && +(2m_\pi^2+q^2)\big[ 18\Gamma_2+6p^2(\Gamma_3-\Gamma_0) +q^2 \Gamma_0\big] +\Big(m_\pi^2+{q^2\over 2}\Big)^{\!2}\big[(6p^2-q^2)G_0-6G_{0*}\big]  \nonumber \\ && +(\vec \sigma_1\!\cdot\!\vec \sigma_2 q^2-\vec \sigma_1\!\cdot\! \vec q\, \vec \sigma_2\!\cdot\! \vec q\,)\big[(2m_\pi^2+q^2)G_2-2\Gamma_2\big]
\nonumber \\ && +\vec\sigma_1\!\cdot\!(\vec q\! \times\! \vec p\,)\vec \sigma_2\!\cdot\!(\vec q\!\times\!\vec p\,)\Big[(2m_\pi^2+ q^2)(G_0+4G_1+4G_3)-2\Gamma_0 -4\Gamma_1 -2\Gamma_3 \Big] \nonumber \\ && +i( \vec\sigma_1\!+\!\vec\sigma_2)\!\cdot\!(\vec q\!\times \! \vec p\,)\bigg[ p^2(4\Gamma_0+5\Gamma_1-2\Gamma_3- 3\Gamma_5) -{q^2\over 2}(3\Gamma_0+ 4\Gamma_1+\Gamma_3)-3 (2\Gamma_2+5\Gamma_4)\nonumber \\ && -m_\pi^2(\Gamma_0+2\Gamma_1)+\Big(m_\pi^2+{q^2\over 2}\Big)\Big(
(m_\pi^2-3p^2+q^2)(G_0+2G_1)+3G_{0*}+6G_{1*}\Big)\bigg]\bigg\}\,, \end{eqnarray}
where the functions $G_{\nu*}$ are defined in the appendix.
\subsection{Matching to original forms}
In this subsection, we analyze how the original forms of the $2\pi$-exchange 3N-interaction \cite{3Nlong,holt} match with the general form modulo shorter range terms introduced in eq.(58). By explicit comparison one finds that at N$^3$LO the differences between the original form in eq.(1) and the represention modulo shorter range terms in eq.(58) are on the one hand accounted for by shifts of the low-energy constants $c_{1,3,4}$:  
\begin{equation}\delta c_1 = {g_A^2 m_\pi\over 64\pi f_\pi^2}\,, \qquad  \delta c_3 = -{g_A^4 m_\pi\over 16\pi f_\pi^2}\,, \qquad  \delta c_4 = {g_A^4 m_\pi\over 16\pi f_\pi^2}\,, \qquad  \end{equation} 
which are exactly opposite to those in eq.(3). On the other hand one has to shift the two functions $f_{6,7}(s)$ entering the $2\pi1\pi$-exchange 3N-interaction (see eq.(11) by: 
\begin{equation}\delta f_6(s) = \delta f_7(s)= (4g_A^2-2)m_\pi-(4m_\pi^2+2s^2)A(s)\,.  \end{equation} 
At N$^2$LO differences occur only with respect to the $c_3$-term. The (total) difference for $V_\text{med}$ between the results derived in ref.\,\cite{holt} and those based on $\tilde g_+(q_2) = c_3(2m_\pi^2+q_2^2)$ reads:
\begin{equation}\delta V^{(c_3)}_\text{med} = {g_A^2 c_3\over 8\pi^2f_\pi^4} \bigg\{ \vec\tau_1\!\cdot\!\vec\tau_2\,\vec\sigma_1\!\cdot\!\vec q\, \vec\sigma_2\!\cdot\!\vec q\,\bigg({2k_f^3\over m_\pi^2+q^2}-\widetilde\Gamma_1\bigg)-4k_f^3+6m_\pi^2 \Gamma_0+3q^2\widetilde\Gamma_1-3 i( \vec\sigma_1\!+\!\vec\sigma_2)\!\cdot\!(\vec q\!\times \! \vec p\,)\widetilde\Gamma_1\bigg\}\,.\end{equation}
Only the first piece proportional to $2k_f^3/(m_\pi^2+q^2)$ is equivalent to a shift of the parameter $c_D$ belonging to the chiral $1\pi$-exchange 3N-force by $\delta c_D= -3c_3 g_A\Lambda_\chi$. The other terms lead to disagreeing  partial-wave matrix elements. As a matter of fact, the decomposition into $2\pi$-exchange and $1\pi$-exchange components, which one can infer from the identity $2c_3 \vec q_1\!\cdot\!\vec q_3=c_3(2m_\pi^2+q_2^2)-c_3(m_\pi^2+q_1^2)-c_3(m_\pi^2+q_3^2)$, is not chiral-invariant. The extra non-chiral-invariant $1\pi$-exchange 3N-interaction 
\begin{equation} \delta V_\text{3N}^{(c_3)} = -{g_A^2c_3 \over 4f_\pi^4} \,{\vec\sigma_3\!\cdot\!\vec q_3 \over m_\pi^2+q_3^2} \, \vec\tau_1\!\cdot\!\vec\tau_3\, \vec \sigma_1\!\cdot\!\vec q_1\,,\end{equation} 
needs to be included, if one works with  $\tilde g_+(q_2) = c_3(2m_\pi^2+q_2^2)$, and it reproduces $\delta V^{(c_3)}_\text{med}$ in eq.(75). The same detailed analysis for the chiral 3N-interaction at fourth order \cite{twopi4,midrange4} is a subject for future work. 
\section*{Appendix: Collection of relevant functions}
In this appendix we give the analytical expressions of functions introduced in the main part of our paper. Two particular (combinations of) $H_{j,\nu}$-functions, which appear in eq.(24), read:
\begin{eqnarray} &&-2H_{5,0}(p)-p^2 H_{4,3}(p)-3H_{4,2}(p) = g_A^2\bigg\{\bigg[ {p^4\over 60}-{k_f^4\over 4}+ 4m_\pi^4 -{k_f^2 p^2\over 6}-{k_f^3 p\over 3}-{k_f^5 \over 15 p}\bigg] \arctan{p+k_f \over 2m_\pi}\nonumber \\ &&\qquad \qquad \quad  + \bigg[{k_f^4\over 4}- {p^4 \over 60}- 4m_\pi^4 +{k_f^2 p^2\over 6}-{k_f^3 p\over 3}-{k_f^5 \over 15 p}\bigg] \arctan{p-k_f \over 2m_\pi}\nonumber \\ && \qquad\qquad  \quad +{2m_\pi^3\over p}\bigg({p^2-k_f^2\over 3}-{4m_\pi^2\over 5}\bigg)\ln{4m_\pi^2+(p+k_f)^2\over 4m_\pi^2+(p-k_f)^2}+{m_\pi k_f\over 5}\Big(3 k_f^2-{p^2\over 3}-12m_\pi^2\Big)\bigg\}\,,  \end{eqnarray} 
\begin{eqnarray}H_{4,1}(p) &=& g_A^2\bigg\{{1\over 3}\bigg[ {p^2\over 5}-k_f^2-{k_f^3 \over p}+{k_f^5 \over 5 p^3}\bigg] \arctan{p+k_f \over 2m_\pi}+{1\over 3}\bigg[ k_f^2-{p^2\over 5}-{k_f^3 \over p}+{k_f^5 \over 5 p^3}\bigg] \arctan{p-k_f \over 2m_\pi} \nonumber \\ &&  +{m_\pi\over p}\bigg[{k_f^2\over 4}- {m_\pi^2\over 3}-{p^2 \over 8}-{k_f^4\over 8p^2}-{k_f^2m_\pi^2\over 3p^2}-{2m_\pi^4\over 5p^2}
\bigg]\ln{4m_\pi^2+(p+k_f)^2\over 4m_\pi^2+(p-k_f)^2} \nonumber \\ && +{m_\pi k_f\over 5p^2}\Big[{7\over 6}(k_f^2+p^2)+2m_\pi^2\Big]\bigg\}\,.  \end{eqnarray}
The other $H_{j,\nu}$ have a similar form, but involve more terms due to both prefactors $1$ and $g_A^2$. 

In the evaluation of the concatenations of two nucleon-lines in subsections 4.2 and 4.3 the following $l$-dependent ($\Gamma,\gamma$)-functions have emerged:  
\begin{eqnarray} \widetilde\Gamma_1(l) &=& {k_f\over 4l^2}(m_\pi^2+k_f^2+l^2) -
                                           {1\over 16l^3}\big[m_\pi^2+(k_f+l)^2\big]\big[m_\pi^2+(k_f-l)^2\big] \ln{m_\pi^2+(k_f+l)^2\over m_\pi^2+(k_f-l)^2}\,,  \end{eqnarray} \begin{eqnarray} \Gamma_2(l) &=& {m_\pi^3 \over 3}\bigg[ \arctan{k_f+l\over m_\pi}+ \arctan{k_f-l\over m_\pi}\bigg]+{k_f\over 9}(k_f^2-3m_\pi^2) + {k_f\over 24l^2}(k_f^2+m_\pi^2)^2 \nonumber \\ &&-{k_f l^2\over 24}+{l^2-k_f^2-m_\pi^2\over 96l^3} \Big[(m_\pi^2+k_f^2)^2+l^4+2l^2(5m_\pi^2-k_f^2)\Big] \ln{m_\pi^2+(k_f+l)^2\over m_\pi^2+(k_f-l)^2}\,,\end{eqnarray} \begin{eqnarray}   \widetilde\Gamma_3(l) &=& {k_f\over 8}+{k_f^3 \over 3l^2}- {k_f\over 8l^4}(k_f^2+m_\pi^2)^2\nonumber \\ && +{m_\pi^2+k_f^2-l^2 \over 32l^5}\big[ m_\pi^2+(k_f+l)^2\big]\big[m_\pi^2+(k_f-l)^2\big] \ln{m_\pi^2+(k_f+l)^2\over m_\pi^2+(k_f-l)^2}  \,, \end{eqnarray} \begin{eqnarray}  \gamma_2(l) &=& 
-{m_\pi \over 2}\bigg[ \arctan{k_f+l\over m_\pi}+ \arctan{k_f-l\over m_\pi}\bigg]
+{k_f\over 8l^2}(3l^2-k_f^2-m_\pi^2)\nonumber \\ && +{1\over 32l^3 } \Big[(m_\pi^2+ k_f^2)^2-3l^4 +2l^2 (3m_\pi^2+k_f^2)\Big] \ln{m_\pi^2+(k_f+l)^2\over  m_\pi^2+(k_f-l)^2}\,, \end{eqnarray} \begin{eqnarray} \widetilde\gamma_3(l) &=& {k_f\over 8l^4}(3m_\pi^2+3k_f^2-l^2)
 +{1\over 32l^5}\Big[l^4+2l^2(k_f^2-m_\pi^2)-3(m_\pi^2+k_f^2)^2\Big] \ln{m_\pi^2+(k_f+l)^2\over m_\pi^2+(k_f-l)^2}\,.  \end{eqnarray} 

The functions $G_\nu$ and $G_{\nu*}$ arising from Fermi sphere integrals $(2\pi)^{-1}\!\int\!d^3l\, \theta(k_f-|\vec l\,|)$ over the product of two different pion-propagator $[m_\pi^2+(\vec l+\vec p\,)^2]^{-1}[m_\pi^2+(\vec l+\vec p\,')^2]^{-1}$, supplemented by tensorial factors $1\,(\nu=0)$, $l_i\,(\nu=1)$, or $l_il_j\,(\nu=2,3)$ have the following representations as (one-parameter) radial integrals: 
\begin{eqnarray} && \{G_0,G_{0*}\}=2\int_0^{k_f}\!\! dl\{l,l^3\} \Omega(l)\,, \\ 
&&   \{G_1,G_{1*}\} = {2\over 4p^2-q^2}  \int_0^{k_f}\!\! dl\{l,l^3\}\big[\Lambda(l)-(m_\pi^2+l^2+p^2) \Omega(l)\big]\,, \\&&   \{G_2,G_{2*}\} = {2\over 4p^2-q^2}  \int_0^{k_f}\!\! dl\{l,l^3\}\big[(m_\pi^2+l^2+p^2)\Lambda(l)-(B+q^2l^2) \Omega(l)\big]\,, \\&&   \{G_3,G_{3*}\} = {1\over 4p^2-q^2}  \int_0^{k_f}\!\! dl\{l,l^3\}\bigg[{l\over 2p^2}-(m_\pi^2+l^2+p^2)\Big({1\over 2p^2}+{4\over  4p^2-q^2} \Big)\Lambda(l)\nonumber \\ && \qquad\qquad\qquad\qquad\qquad\quad\qquad\qquad +\Big({4(m_\pi^2+l^2+p^2)^2 \over 4p^2-q^2}-2l^2\Big)\Omega(l)\bigg]\,, \end{eqnarray}  with $\Lambda(l), \Omega(l)$ and $B$ defined in eqs.(42,43) and below. Note that a $*$ designates an additional power of $l^2$ in the integrand.
\section*{Acknowledgements}
We thank E. Epelbaum and H. Krebs for detailed information about chiral three-nucleon forces. The work of B. Singh has been supported by a Professional Enhancement Grant of the Tata Trusts, India.

\end{document}